\documentclass[conference]{IEEEtran}

\usepackage[sort, nospace]{cite}
\usepackage{amsmath,amssymb,amsfonts}
\usepackage{algorithmic}
\usepackage{graphicx}
\usepackage{textcomp}
\usepackage{xcolor}
\usepackage[hyphens]{url}

\usepackage{xspace}
\usepackage{balance}
\usepackage{enumitem}
\usepackage{multirow}
\usepackage{mathrsfs}
\usepackage{amsthm}
\usepackage{setspace}
\usepackage{authblk}
\usepackage{url}
\usepackage{array}
\usepackage{tabularx, booktabs}
\usepackage{float}
\usepackage{soul}
\usepackage{mathtools}
\usepackage{braket}
\usepackage{tikz}
\usepackage{pgfplots}
\usepackage{xcolor}
\usepackage{graphicx}
\usepackage{epsfig}
\usepackage{pgfmath}
\usepackage{savesym}
\usepackage{tabulary}
\usepackage{booktabs}
\usepackage{multirow}
\usepackage{caption}
\usepackage{subcaption}
\usepackage{algorithm}
\usepackage{algorithmic}
\usepackage{qcircuit}
\usepackage{hyperref}

\usepackage{authblk}
\usetikzlibrary{positioning, calc, shapes, arrows}
\usetikzlibrary{decorations.pathreplacing, calligraphy, backgrounds}
\usetikzlibrary{fit}
\def\BibTeX{{\rm B\kern-.05em{\sc i\kern-.025em b}\kern-.08em
    T\kern-.1667em\lower.7ex\hbox{E}\kern-.125emX}}

\title{Mitigating Coupling Map Constrained Correlated Measurement Errors on Quantum Devices} 
\author[1]{Alan Robertson\thanks{alan.robertson@sydney.edu.au}}
\author[2]{Shuaiwen Leon Song\thanks{leon.song@university.edu.au}}
\affil[1, 2]{School of Computer Science, University of Sydney}
\renewcommand{\baselinestretch}{.97}

\begin{document}
\maketitle
\thispagestyle{plain}
\pagestyle{plain}

%%%%%% -- PAPER CONTENT STARTS-- %%%%%%%%

\begin{abstract}
We introduce a technique for the suppression of state-dependent and correlated measurement errors, which are commonly observed on modern superconducting quantum devices. Our method leverages previous results, establishing that correlated errors tend to be physically localised on quantum devices to perform characterisations over the coupling map of the device, and to join overlapping measurement calibrations as a series of sparse matrices. We term this `Coupling Map Calibration'. We quantitatively demonstrate the advantages of our proposed error mitigation system design across a range of current IBM quantum devices. Our experimental results on common benchmark circuits demonstrate up to a $41\%$ reduction in the error rate without increasing the number of executions of the quantum device required when compared to conventional error mitigation methods.

%\textcolor{blue}{We demonstrate the scalability of this technique for a range of common NISQ device topologies, and its performance compared to other state-of-the-art techniques.}
\end{abstract}
\section{Introduction}

Advancements in techniques to accurately store and process quantum information have led towards the realisation of small scale prototype quantum devices~\cite{preskill_nisq,Qiskit,sycamore,honeywell,ionq,rigetti}. It is hoped that these near term, intermediate scale quantum devices (NISQ) may improve to the point that they are able to implement algorithms with asymptotic improvements over their classical counterparts~\cite{deutsch_rapid_1992, shor_polynomial-time_1995, grover_fast_1996}. These algorithms present more efficient methods for tackling problems such as integer factorisation and nano-scale simulation~\cite{kassal_polynomial-time_2008,full_stack}. 

One of a range of hurdles that hinders the potential applications of quantum computation is the extreme sensitivity to noise~\cite{nielsen} and the associated difficulty in ensuring the reliable execution of quantum algorithms. Error rates are too high and qubit counts too low to perform quantum error correction and satisfy a fault tolerance threshold~\cite{Proctor_2021,gottesman}. In this noisy regime, devices are limited to low circuit depths and require a large number of repeated executions of a quantum circuit to statistically determine the correct output~\cite{altepeter}. One of the major challenges in designing future quantum systems is to mitigate or even eliminate the sources of errors with the hope of eventually achieving fault tolerance and scalable computation.

Several approaches have been taken to the problems of characterising, suppressing\cite{magesan_characterizing_2012,emerson_rb,altepeter,Granade_2017}, mitigating \cite{harper_correlated,howard,martonosi_crosstalk}, and correcting errors\cite{gottesman} in quantum systems. As quantum devices require regular recalibration, long-term characterisation is not always possible as the exhibition of errors on these systems drifts over time~\cite{harper_fault-tolerant_2019}. This characterisation is essential for mitigating errors in quantum systems. Once characterised, a variety of techniques may be deployed to attempt to suppress the associated errors, and in the long term approach a fault tolerant threshold.

\begin{figure}[t!]
\begin{subfigure}[b]{0.15\textwidth}
\centering
\begin{tikzpicture}
\def \radius {0.7cm}
\node[draw, circle, violet!40!pink!40!black!40, fill=blue!60!black!40, text=white] (n_0) {$0$};
\node[draw, circle, violet!40!pink!40!black!40, fill=blue!60!black!40, text=white, right of = n_0, yshift=-0.75em]  (n_1) {$1$};
\node[draw, circle, violet!40!pink!40!black!40, fill=blue!60!black!40, text=white, right of = n_1, yshift=0.75em]  (n_2) {$2$};
\node[draw, circle, violet!40!pink!40!black!40, fill=blue!60!black!40, text=white, below of = n_1, yshift=0.75em]  (n_3) {$3$};
\node[draw, circle, violet!40!pink!40!black!40, fill=blue!60!black!40, text=white, below of = n_3, yshift=0.75em]  (n_5) {$5$};
\node[draw, circle, violet!40!pink!40!black!40, fill=blue!60!black!40, text=white, left of = n_5, yshift = -0.75em]  (n_4) {$4$};
\node[draw, circle, violet!40!pink!40!black!40, fill=blue!60!black!40, text=white, right of = n_5, yshift = -0.75em]  (n_6) {$6$};

\draw [violet!77!black!23,line width=0.48899999999999993mm,-] (n_0) -- (n_1);
\draw [violet!74!black!26,line width=0.549mm,-] (n_0) to [bend left=15] (n_2);
\draw [violet!69!black!31,line width=0.669mm,-] (n_0) -- (n_3);
\draw [violet!73!black!27,line width=0.5820000000000001mm,-] (n_0) to [bend left=15] (n_4);
\draw [violet!72!black!28,line width=0.603mm,-] (n_0) -- (n_5);
%\draw [violet!69!black!31,line width=0.669mm,-] (n_0) -- (n_6);
\draw [violet!83!black!17,line width=0.372mm,-] (n_1) -- (n_2);
\draw [violet!20!black!80,line width=2.19mm,-] (n_1) -- (n_3);
\draw [violet!75!black!25,line width=0.534mm,-] (n_1) -- (n_4);
\draw [violet!78!black!22,line width=0.471mm,-] (n_1) to [bend left=60] (n_5);
\draw [violet!73!black!27,line width=0.5790000000000001mm,-] (n_1) -- (n_6);
\draw [violet!78!black!22,line width=0.48mm,-] (n_2) -- (n_3);
%\draw [violet!77!black!23,line width=0.49200000000000005mm,-] (n_2) -- (n_4);
\draw [violet!78!black!22,line width=0.46499999999999997mm,-] (n_2) -- (n_5);
\draw [violet!74!black!26,line width=0.5579999999999999mm,-] (n_2) to [bend right=15] (n_6);
\draw [violet!65!black!35,line width=0.747mm,-] (n_3) -- (n_4);
\draw [violet!67!black!33,line width=0.7140000000000001mm,-] (n_3) -- (n_5);
\draw [violet!71!black!29,line width=0.633mm,-] (n_3) -- (n_6);
\draw [violet!71!black!29,line width=0.627mm,-] (n_4) -- (n_5);
\draw [violet!66!black!34,line width=0.7320000000000001mm,-] (n_4) to [bend right=15] (n_6);
\draw [violet!29!black!71,line width=1.536mm,-] (n_5) -- (n_6);

\end{tikzpicture}
\caption{\scriptsize IBMQ Oslo\label{subfig:corr_oslo}}
\end{subfigure}
\hfill
\begin{subfigure}[b]{0.15\textwidth}
%\centering
\begin{tikzpicture}
\def \radius {0.9cm}
\node[draw, circle, violet!40!black!40, fill=blue!60!black!40, text=white] (n_0) {$0$};
\node[draw, circle, violet!40!black!40, fill=blue!60!black!40, text=white, right of = n_0, yshift = -0.75em] (n_1) {$1$};
\node[draw, circle, violet!40!black!40, fill=blue!60!black!40, text=white, right of = n_1, yshift = 0.75em]  (n_2) {$2$};
\node[draw, circle, violet!40!black!40, fill=blue!60!black!40, text=white, below of = n_1, yshift = 0.35em] (n_3) {$3$};
\node[draw, circle, violet!40!black!40, fill=blue!60!black!40, text=white, below of = n_3, yshift = 0.35em]  (n_4) {$4$};

\draw [violet!36!black!64,line width=0.46799999999999997mm,-] (n_0) -- (n_1);
\draw [violet!20!black!80,line width=0.738mm,-] (n_0) to [bend left=15] (n_2);
\draw [violet!20!black!80,line width=0.687mm,-] (n_0) -- (n_3);
\draw [violet!20!black!80,line width=0.705mm,-] (n_0) -- (n_4);
\draw [violet!46!black!54,line width=0.396mm,-] (n_1) -- (n_2);
\draw [violet!28!black!72,line width=0.525mm,-] (n_1) -- (n_3);
\draw [violet!32!black!68,line width=0.498mm,-] (n_1) to [bend left=60] (n_4);
\draw [violet!24!black!76,line width=0.5549999999999999mm,-] (n_2) -- (n_3);
\draw [violet!22!black!78,line width=0.573mm,-] (n_2) -- (n_4);
\draw [violet!20!black!80,line width=0.591mm,-] (n_3) -- (n_4);
\end{tikzpicture}
\caption{\scriptsize IBMQ Lima\label{subfig:corr_lima}}
\end{subfigure}
\hfill
\begin{subfigure}[b]{0.15\textwidth}
\centering
\begin{tikzpicture}
\def \radius {0.9cm}
\node[draw, circle, violet!40!black!40, fill=blue!60!black!40, text=white] (n_0) {$0$};
\node[draw, circle, violet!40!black!40, fill=blue!60!black!40, text=white, right of = n_0, yshift = -0.75em] (n_1) {$1$};
\node[draw, circle, violet!40!black!40, fill=blue!60!black!40, text=white, right of = n_1, yshift = 0.75em]  (n_2) {$2$};
\node[draw, circle, violet!40!black!40, fill=blue!60!black!40, text=white, below of = n_1, yshift = 0.35em] (n_3) {$3$};
\node[draw, circle, violet!40!black!40, fill=blue!60!black!40, text=white, below of = n_3, yshift = 0.35em]  (n_4) {$4$};

\draw [violet!76!black!24,line width=0.47400000000000003mm,-] (n_0) -- (n_1);
\draw [violet!74!black!26,line width=0.513mm,-] (n_0) to [bend left=15] (n_2);
\draw [violet!55!black!45,line width=0.915mm,-] (n_0) -- (n_3);
\draw [violet!33!black!67,line width=1.353mm,-] (n_0) -- (n_4);
\draw [violet!68!black!32,line width=0.642mm,-] (n_1) -- (n_2);
\draw [violet!54!black!46,line width=0.9299999999999999mm,-] (n_1) -- (n_3);
\draw [violet!36!black!64,line width=1.293mm,-] (n_1) to [bend left=60] (n_4);
\draw [violet!50!black!50,line width=1.0110000000000001mm,-] (n_2) -- (n_3);
\draw [violet!34!black!66,line width=1.332mm,-] (n_2) -- (n_4);
\draw [violet!20!black!80,line width=2.0429999999999997mm,-] (n_3) -- (n_4);

\end{tikzpicture}
\caption{\scriptsize  IBMQ Quito\label{subfig:corr_quito}}
\end{subfigure}
\hfill
\begin{subfigure}[b]{0.15\textwidth}
\centering
\begin{tikzpicture}
\def \radius {0.9cm}
\node[draw, circle, violet!40!pink!40!black!40, fill=blue!60!black!40, text=white] at ({360/5 * (1) + 20}:\radius) (n_0) {$0$};
\node[draw, circle, violet!40!pink!40!black!40, fill=blue!60!black!40, text=white] at ({360/5 * (2) + 20}:\radius) (n_1) {$1$};
\node[draw, circle, violet!40!pink!40!black!40, fill=blue!60!black!40, text=white] at ({360/5 * (3) + 20}:\radius) (n_2) {$2$};
\node[draw, circle, violet!40!pink!40!black!40, fill=blue!60!black!40, text=white] at ({360/5 * (4) + 20}:\radius) (n_3) {$3$};
\node[draw, circle, violet!40!pink!40!black!40, fill=blue!60!black!40, text=white] at ({360/5 * (5) + 20}:\radius) (n_4) {$4$};

\draw [violet!89!black!11,line width=0.375mm,-] (n_0) -- (n_1);
\draw [violet!87!black!13,line width=0.432mm,-] (n_0) -- (n_2);
\draw [violet!85!black!15,line width=0.516mm,-] (n_0) -- (n_3);
\draw [violet!86!black!14,line width=0.471mm,-] (n_0) -- (n_4);
\draw [violet!32!black!68,line width=1.173mm,-] (n_1) -- (n_2);
\draw [violet!20!black!80,line width=1.7355mm,-] (n_1) -- (n_3);
\draw [violet!20!black!80,line width=1.7085mm,-] (n_1) -- (n_4);
\draw [violet!89!black!11,line width=0.372mm,-] (n_2) -- (n_3);
\draw [violet!89!black!11,line width=0.348mm,-] (n_2) -- (n_4);
\draw [violet!91!black!9,line width=0.29700000000000004mm,-] (n_3) -- (n_4);
\end{tikzpicture}
\caption{\scriptsize IBMQ Manila\label{subfig:corr_manila}}
\end{subfigure}
\hfill
\begin{subfigure}[b]{0.15\textwidth}
\centering
\begin{tikzpicture}
\def \radius {0.9cm}
\node[draw, circle, violet!40!pink!40!black!40, fill=blue!60!black!40, text=white] (n_0) {$0$};
\node[draw, circle, violet!40!pink!40!black!40, fill=blue!60!black!40, text=white, right of = n_0, yshift=-0.75em]  (n_1) {$1$};
\node[draw, circle, violet!40!pink!40!black!40, fill=blue!60!black!40, text=white, right of = n_1, yshift=0.75em]  (n_2) {$2$};
\node[draw, circle, violet!40!pink!40!black!40, fill=blue!60!black!40, text=white, below of = n_1, yshift=0.75em]  (n_3) {$3$};
\node[draw, circle, violet!40!pink!40!black!40, fill=blue!60!black!40, text=white, below of = n_3, yshift=0.75em]  (n_5) {$5$};
\node[draw, circle, violet!40!pink!40!black!40, fill=blue!60!black!40, text=white, left of = n_5, yshift = -0.75em]  (n_4) {$4$};
\node[draw, circle, violet!40!pink!40!black!40, fill=blue!60!black!40, text=white, right of = n_5, yshift = -0.75em]  (n_6) {$6$};

\draw [violet!55!black!45,line width=0.63mm,-] (n_0) -- (n_1);
\draw [violet!20!black!80,line width=1.377mm,-] (n_0) to [bend left=15] (n_2);
\draw [violet!48!black!52,line width=0.726mm,-] (n_0) -- (n_3);
\draw [violet!70!black!30,line width=0.408mm,-] (n_0) to [bend left=15] (n_4);
\draw [violet!69!black!31,line width=0.435mm,-] (n_0) -- (n_5);
%\draw [violet!59!black!41,line width=0.573mm,-] (n_0) -- (n_6);
\draw [violet!55!black!45,line width=0.618mm,-] (n_1) -- (n_2);
\draw [violet!38!black!62,line width=0.864mm,-] (n_1) -- (n_3);
\draw [violet!64!black!36,line width=0.495mm,-] (n_1) -- (n_4);
\draw [violet!61!black!39,line width=0.543mm,-] (n_1) to [bend left=60] (n_5);
\draw [violet!61!black!39,line width=0.5399999999999999mm,-] (n_1) -- (n_6);
\draw [violet!41!black!59,line width=0.8220000000000001mm,-] (n_2) -- (n_3);
%\draw [violet!67!black!33,line width=0.462mm,-] (n_2) -- (n_4);
\draw [violet!66!black!34,line width=0.462mm,-] (n_2) -- (n_5);
\draw [violet!69!black!31,line width=0.423mm,-] (n_2) to [bend right=15] (n_6);
\draw [violet!37!black!63,line width=0.873mm,-] (n_3) -- (n_4);
\draw [violet!20!black!80,line width=1.4040000000000001mm,-] (n_3) -- (n_5);
\draw [violet!33!black!67,line width=0.9299999999999999mm,-] (n_3) -- (n_6);
\draw [violet!67!black!33,line width=0.453mm,-] (n_4) -- (n_5);
\draw [violet!20!black!80,line width=1.401mm,-] (n_4) to [bend right=15] (n_6);
\draw [violet!61!black!39,line width=0.543mm,-] (n_5) -- (n_6);
\end{tikzpicture}
\caption{\scriptsize  IBMQ Nairobi\label{subfig:corr_nairobi}}
\end{subfigure}
\hfill
\begin{subfigure}[b]{0.15\textwidth}
\centering
\begin{tikzpicture}
\def \radius {0.9cm}
\node[draw, circle, violet!40!black!40, fill=blue!60!black!40, text=white] (n_0) {$0$};
\node[draw, circle, violet!40!black!40, fill=blue!60!black!40, text=white, right of = n_0, yshift = -0.75em] (n_1) {$1$};
\node[draw, circle, violet!40!black!40, fill=blue!60!black!40, text=white, right of = n_1, yshift = 0.75em]  (n_2) {$2$};
\node[draw, circle, violet!40!black!40, fill=blue!60!black!40, text=white, below of = n_1, yshift = 0.35em] (n_3) {$3$};
\node[draw, circle, violet!40!black!40, fill=blue!60!black!40, text=white, below of = n_3, yshift = 0.35em]  (n_4) {$4$};

\draw [violet!60!black!40,line width=0.402mm,-] (n_0) -- (n_1);
\draw [violet!54!black!46,line width=0.462mm,-] (n_0) to [bend left=15] (n_2);
\draw [violet!46!black!54,line width=0.546mm,-] (n_0) -- (n_3);
\draw [violet!20!black!80,line width=0.8099999999999999mm,-] (n_0) -- (n_4);
\draw [violet!46!black!54,line width=0.549mm,-] (n_1) -- (n_2);
\draw [violet!47!black!53,line width=0.5399999999999999mm,-] (n_1) -- (n_3);
\draw [violet!20!black!80,line width=1.026mm,in=135,out=225] (n_1) to [bend left=60] (n_4);
\draw [violet!44!black!56,line width=0.5640000000000001mm,-] (n_2) -- (n_3);
\draw [violet!49!black!51,line width=0.513mm,-] (n_2) -- (n_4);
\draw [violet!40!black!60,line width=0.615mm,-] (n_3) -- (n_4);
\end{tikzpicture}
\caption{\scriptsize  IBMQ Belem\label{subfig:corr_belem}}
\end{subfigure}
\caption{\small Frobenius norm between calibration matrices $C_{ij}$ and $C_{i} \otimes C_{j}$ for all pairs of qubits over a range of IBM quantum devices averaged over three weeks. Thicker edges indicate a greater correlation of measurement errors between those qubits.\label{fig:corr_coupling_demo}}
\end{figure}
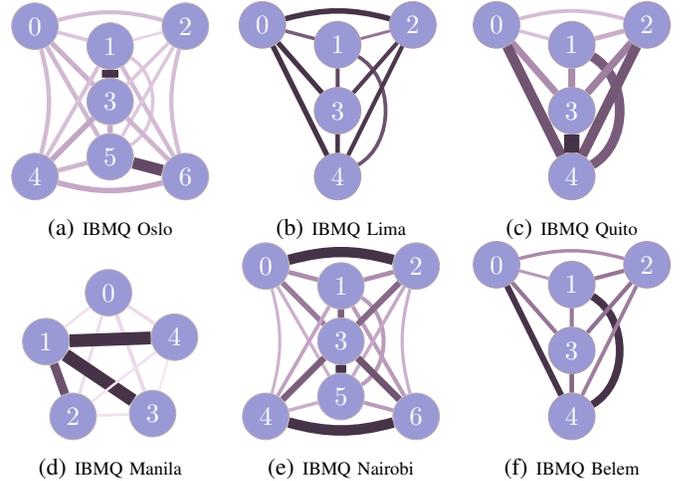

 Current state-of-the-art characterisation techniques include randomised benchmarking~\cite{magesan_characterizing_2012}, tomography~\cite{berry,howard,altepeter}, heuristic methods~\cite{Granade_2017}, and particle filtering techniques~\cite{riddhi, qinfer-1_0}. In this paper, we focus on suppressing two categories of significant errors widely observed on today's superconducting NISQ devices: \textit{state-dependent measurement errors} and \textit{correlated measurement errors}. Prior approaches~\cite{swamit_state_dep,Qiskit} to suppress these errors have been hemmed in by the competing goals of scalability and complexity of characterising correlated errors. More recent Bayesian approaches~\cite{jigsaw} are constrained by sampling requirements on devices that exhibit inherently correlated, non-uniform and non-Markovian noise~\cite{wildcard_rbk}. An example of the complexity of this error landscape can be seen in Fig.~\ref{fig:corr_coupling_demo}. Here the physical layout of the qubits is represented by their proximity in the figure, while the correlation of measurement errors is represented by the edge thickness.% of the edges between the qubits.

To address this challenge, we propose a {\it Coupling Map Calibration (CMC)} scheme, which represents a middle ground approach that efficiently stitches together small measurement error calibration patches and preserves the information about local correlations. To extend this approach, we also propose a coupling map profiling method {\it ERR} which constructs maps of the highest locally-correlated qubits measurements which may then be passed to CMC. In summary, this paper makes the following contributions: 

\begin{itemize}
%\item We verify the continuing existence of state-dependent and correlated measurement errors and correlated errors on current IBMQ devices;
\item We demonstrate a method for efficiently constructing and joining sparse measurement calibration matrices to form a full-device measurement calibration matrix. From this we construct CMC: a scheme for performing joint sparse calibrations using the coupling maps of NISQ devices;
\item We extend this method to account for local but non-coupling map aligned error models with a patching technique termed {\it ERR}. We demonstrate that ERR maps are stable on the order of several weeks;
\item Our experimental results in implementing CMC and CMC-ERR on a range of IBMQ devices reduce errors by an average of $35\%$ and up to $41\%$, equalling or outperforming existing scalable methods. We additionally demonstrate that these results are consistent with expectations of measurement errors from simulation. We additionally demonstrate that the choice of measurement error mitigation technique depends upon the noise-profile and architectural topologies of the device.
\end{itemize}
	
% \end{itemize} 

\section{Background}		
\label{section:background}
\subsection{Quantum Computation with Qubits}
The basis for classical binary computation is the manipulation of a set of discrete two-level systems with states labelled $0$ or $1$: \emph{bits}. A classical computer with $n$ bits can access $2^{n}$ states represented by a string of $0$s and $1$s. By comparison, quantum computation is the manipulation of a set of continuous systems, the most commonly proposed of which is a similar ensemble of two-level systems, termed \emph{qubits}. The states of these quantum systems are labelled using the basis vectors $\ket{0} = \left[\begin{smallmatrix} 1  \\ 0 \end{smallmatrix}\right]$ and $\ket{1} = \left[\begin{smallmatrix} 0  \\ 1 \end{smallmatrix}\right]$. In addition to these basis states, the system may access a series of intermediary linear combinations of states denoted by $\ket{\psi} = \alpha \ket{0} + \beta \ket{1}$. Here $\ket{\psi}$ is an arbitrary qubit and $\alpha$ and $\beta$ are complex numbers such that  $\left|\alpha\right|^2 + \left|\beta\right|^2 = 1$. This linear combination of states is not replicated in classical computation. As $\alpha$ and $\beta$ are normalised complex numbers we may define a set of orthogonal bases $\hat{x}, \hat{y}$ and $\hat{z}$. For a particular choice of measurement basis $\left|\alpha\right|^2$ and $|\beta|^2$ describe the probability of measuring their associated state. We can extend this to a density state representation, where $\rho = \sum_j p_j \ket{\psi_j}\bra{\psi_j}$ and operators are performed by conjugation. Measurements of density states are performed by tracing over a measurement operator $M$. The choice of these measurement operators will be important for our model.

Operations on quantum systems (termed {\it gates}) are unitary operations that act on the state and alter the observed likelihood distribution of measurement outcomes. The general rotation of a single qubit by angles $\theta$ about the X axis, $\phi$ about the Y axis and $\lambda$ about the Z axis, can be described by
\begin{equation}
U3(\theta, \phi, \lambda) = \begin{pmatrix} \cos(\frac{\theta}{2}) & -e^{i\lambda}\sin(\frac{\theta}{2}) \\ e^{i\phi}\sin(\frac{\theta}{2}) & e^{i(\phi+\lambda)}\cos(\frac{\theta}{2}) \end{pmatrix} \label{equ:unitary}.
\end{equation}
The gates $RX(\theta)$ $RZ(\lambda)$ and $RY(\phi)$ are then special cases of this unitary gate and constitute the {\it Pauli Generators}. A rotation by each of these generators by an angle of $\pi$ then gives the $X$, $Y$ and $Z$ gates. While multi-qubit systems can be constructed using the tensor product of two single qubit systems, not all quantum states or operations may be described as a linear combination of single-qubit systems. Some useful multi-qubit operations include invertible controlled gates that conditionally apply a-single qubit rotation to a target qubit. It is possible to deconstruct an arbitrary quantum operation into a sequence of single qubit rotations and two-qubit controlled operations~\cite{nielsen}. A sequence of quantum gates is termed a {\it circuit} (an example of which may be found in Fig.~\ref{fig:x_gates}). Horizontal lines in a circuit represent qubits, while `boxes' represent gates, the progression from left to right indicates the order in which these gates are applied.

\subsection{Types of Quantum Noise}
Given the probabilistic nature of quantum systems, even a small amount of noise results in a non-zero probability of obtaining an incorrect measurement outcome. To mitigate this quantum circuits are executed multiple times in order to collect statistics and attempt to reconstruct an `error-free' output. Each iteration may be referred to as a `shot' or a `trial'. This repetition adds a multiplicative factor to the complexity of the execution of the quantum circuit. If the error rate is sufficiently high, it may not be possible to distinguish the output distribution from the noise. Errors may be characterised as being due to interactions with the environment~\cite{unruh}, imperfect gate operations or, as we focus on in this paper, due to imperfect state preparation and measurement (SPAM).

\textbf{Gate Errors} describe errors due to imperfect gate operations. As measurement is probabilistic any imperfect gate will result in a non-zero probability of producing an incorrect output. This noise will accumulate as the circuit depth grows.

\textbf{SPAM Errors} relate to the incorrect preparation and measurement of the system. Unlike the environmental and gate errors, they only occur during the preparation and measurement stages of the circuit and do not scale with circuit depth, though they do scale with the size of the quantum register, and the number of measurements required.

\subsection{State-Dependent, Biased and Correlated Errors}
In addition to the causes of errors on quantum devices we may also consider how errors act on quantum states. An example error is an independent and identically distributed (IID) uniform random Pauli error over all qubits. This is the same as applying the operation $E(p) = (1 - p)I + \frac{p}{3}(X + Y + Z)$ independently to each qubit in the system for some error rate $p$. Other types of errors are not as impartial. In this paper we will focus on two significant categories of NISQ errors: state-dependent errors and correlated errors, both of which have been observed on NISQ devices~\cite{swamit_state_dep,harper_correlated,jigsaw}. 

\textbf{State-Dependent} errors exist where the error rate associated with an operation depends on the state of the qubit~\cite{swamit_state_dep} with a particular focus on state-dependent errors that occur during measurement. During the relatively long period of time required for measurement on NISQ superconducting devices, the rate of decay from the $\ket{1}$ state is greater than the rate at which qubits in the $\ket{0}$ state are spontaneously excited.

\textbf{Correlated Errors} occur when the probability of an error acting on two qubits is greater than the product of the probabilities with which the error would act on either qubit individually~\cite{harper_correlated,jigsaw}, as demonstrated in Fig.~\ref{fig:correlation}. As most modern quantum devices only admit two-qubit gates between a subset of the qubits on the device, this network of admissible two-qubit relations forms a {\it coupling map}. It has been observed that correlated errors tend to occur in close proximity on the coupling map\cite{harper_correlated}.

\begin{figure}[h]
\include{circs/correlation}
\caption[]{\small Correlated errors. If the error rate of applying A and B simultaneously is greater than the product of the independent error rates, then a correlated error has occurred. Scheduling may be used to mitigate these errors at the cost of an increased circuit depth\label{fig:correlation}.}
\end{figure}

\subsection{Measurement Errors on IBMQ Devices}

\begin{figure}
\centering
\includegraphics[width=\linewidth]{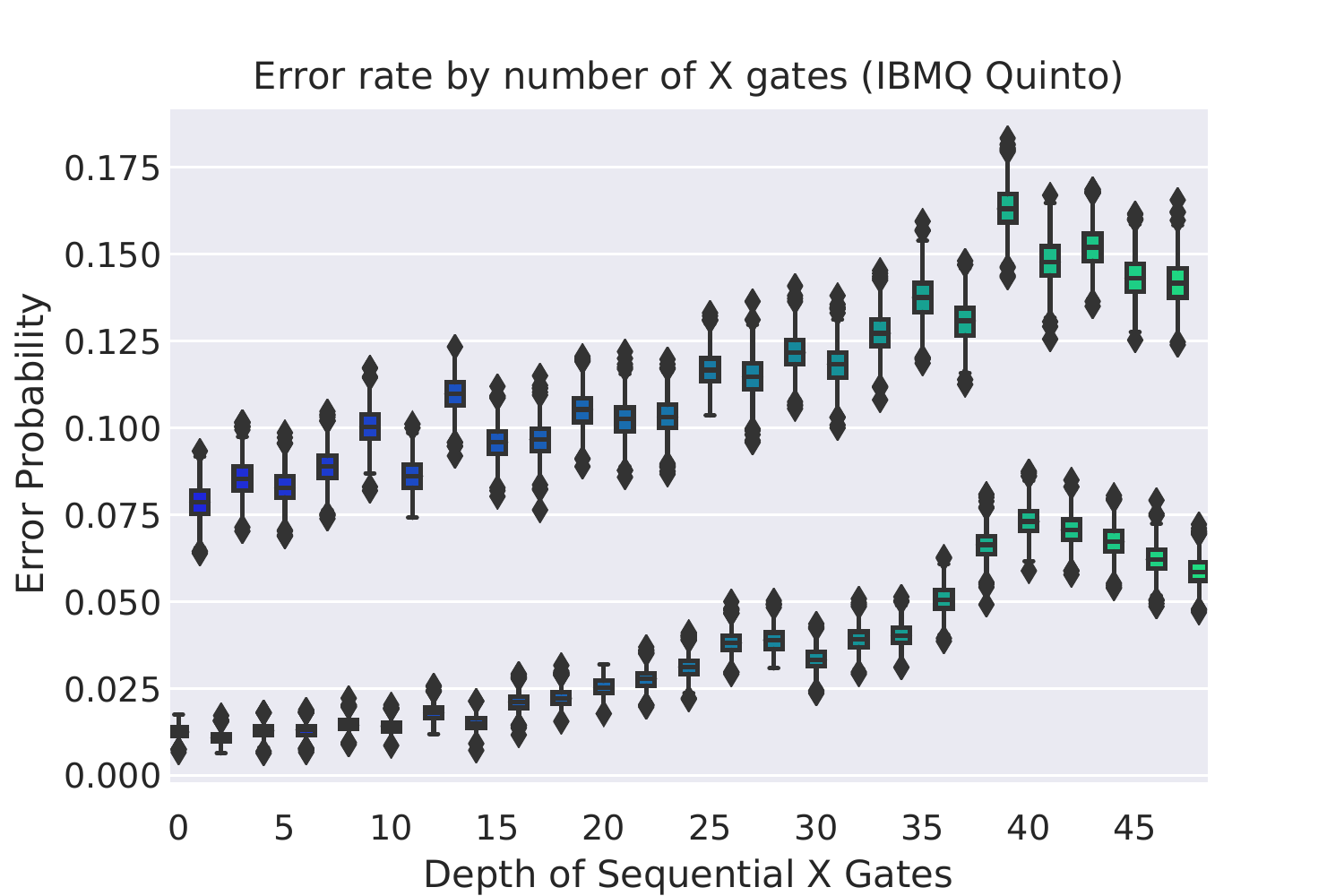}
\includegraphics[width=1.1\linewidth]{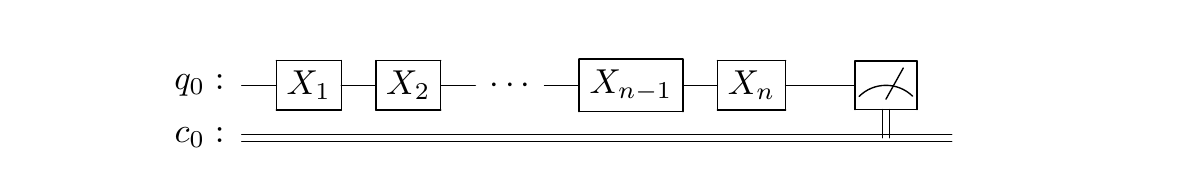}
\caption{\small Error probability over 4000 shots of and Quito device following the application of a sequential number of X gates. Measuring a $\ket{1}$ state exhibits a significantly higher error rate than the $\ket{0}$ state indicating a state-dependent measurement error.}
\label{fig:x_gates}
\end{figure}

In this paper, we focus on state-dependent and correlated measurement errors as currently exhibited on IBMQ devices~\cite{harper_correlated,swamit_state_dep,jigsaw}. State-dependent measurement errors are still present in the most current versions of IBMQ systems as demonstrated in Fig.~\ref{fig:x_gates}. Here a single qubit has been prepared in the $\ket{0}$ state before a sequence of $X$ gates have been applied. As $X$ is a unitary operation, if an odd number of $X$ gates are applied the expected state is $\ket{1}$, while an even number should result in the $\ket{0}$ state. Transpiler optimisations have been disabled to prevent the compiler reducing these gate sequences to either a single $X$ gate or no gate at all.

\begin{table*}[t!]\small
\centering
\begin{tabular}{lcr}
        \toprule
    {\bf Method} & {\bf Computational Cost (Quantum Circuit Executions)} & {\bf Output} \\
    \midrule
    Process Tomography~\cite{howard,altepeter,merkel_tomography} & $r4^{n}$ & SPAM + Gate Errors \\
    Complete Calibration~\cite{Qiskit} & $r2^{n}$ & SPAM Errors \\
    Tensored Calibrations~\cite{Qiskit} & $2nr$ & Non-correlated SPAM Errors \\
    Randomised Benchmarking~\cite{emerson_rb,magesan_characterizing_2012} & $\text{Poly}(n)$ & Average SPAM and Gate \\
    Pauli / Clifford Twirling~\cite{harper_correlated} & $\text{Poly}(n)$ & SPAM-Free Errors \\
    AIM~\cite{swamit_state_dep} & $4r$ & Average Biased SPAM \\
    SIM~\cite{swamit_state_dep} & $2nr + kr$ & Top $k$ least biased SPAM \\
    JIGSAW~\cite{jigsaw} & $\frac{nk}{2}$ + k & Bayesian Error Distribution\\
    {\it CMC (This Paper)} & $\frac{4}{k}er$ & Local SPAM Errors \\
    \bottomrule
\end{tabular}
\caption{\small Computational cost for a range of error characterisation methods. $n$ is the number of qubits and $r$ is the number of repetitions required of the circuit. The $k$ for AIM is a pre-chosen constant, typically 4. For the coupling map method (CMC), $e$ is the number of edges in the graph while $k$ is the speed-up from performing non-local patches simultaneously.\label{table:comparisons}}
\end{table*}

If measurement errors were state independent we would expect that the error rate would increase exponentially with the circuit depth. Instead we observe that the error rate of the $\ket{1}$ states is significantly higher than that of the $\ket{0}$ states even as circuit depth increases, indicating the presence of state-dependent measurement errors. These results suggest that state-dependent measurement errors dominate both single-qubit gate errors and environmental noise at small circuit depths and up to the quantum volume of the device.

Additionally correlated measurement errors are present on current IBMQ devices as seen in Fig.~\ref{fig:corr_coupling_demo}. Here we compare the difference between two single-qubit calibrations $C_i \otimes C_j$ and two-qubit calibrations $C_{ij}$. If the measurement errors were entirely independent then these two calibrations would be equal. This demonstrates that not only do correlated measurement errors exist on these devices, but that some appear to persist between calibration cycles. Previous work has demonstrated that these correlated errors tend to occur in physical proximity on the device~\cite{harper_correlated}.

\section{Measurement Error Suppression}

Previous approaches in error mitigation have involved two often distinct lines of inquiry: the characterisation of the noise of the system~\cite{riddhi,magesan_characterizing_2012,Granade_2017,harper_fault-tolerant_2019}, and the design of the system to reduce noise~\cite{swamit_state_dep, sc_estimating}. Often this may be an iterative approach as errors cannot be suppressed or mitigated until they have been characterised. 

The output from an execution of a quantum system is a single sample from a larger probability distribution. In order to obtain any meaningful statistics to characterise the device, an ensemble of measurements must be taken. Measurement perturbs the state of a quantum system, so a diminishing amount of information is obtained by performing multiple measurements on the same qubit. Cloning a quantum state is also not permitted~\cite{wooters} which prevents ``copying'' a state and performing the same operations on the copies. As a result the most common approach to quantum characterisation is to repeatedly prepare a state and then perform measurements on it~\cite{berry}. The primary figure of merit in comparing approaches to characterising quantum systems is the requisite number of circuits that must be executed. In this section we briefly sketch the details of a range of schemes that seek to mitigate measurement errors on quantum devices. These schemes trade between quantum computational overheads and the scope and quality of the data that is obtained. A comparison of previous characterisation and

\subsection{Tomography} 
One of the first and most accurate methods for characterising quantum states and processes, including errors, is \textit{tomography}~\cite{Granade_2017,altepeter,howard}. As the number of measurements required to perform tomography scales exponentially with the number of qubits, these approaches have become increasingly infeasible on recent devices. Tomography is performed by repeating an experiment over all measurement bases to reconstruct the density state of the system prior to measurement~\cite{greenbaum_tomography,altepeter}. Tomography scales exponentially in the number of qubits but provides the most accurate reconstruction of the state, and by extension the most accurate profile of the noise acting on it~\cite{howard,merkel_tomography}. 

The insight that ties process tomography to error characterisation is that errors are themselves quantum processes that act on the device. By performing process tomography over a portion of the system, environmental, gate and SPAM errors can be diagnosed. i.e. an error is simultaneously an error and an operation that evolves the state and can hence be characterised. Process tomography provides a description of the error channel that incorporates information about state-dependent and correlated errors. Once an error $\mathcal{E}$ is characterised and if the matrix describing $\mathcal{E}$ is invertible, then we can apply $\mathcal{E}^{-1}$ to mitigate the error. The downside is the cost: as the number of qubits increases, repetitions of $4^n$ circuits become computationally unfeasible for modern quantum devices, let alone future devices with hundreds to thousands of qubits.  
The reconstruction of a quantum state from a set of measurements is termed {\it quantum state tomography}~\cite{howard, altepeter}. By taking a histogram of the measurement results over a complete basis of $2^n$ measurement operators, the resulting probability distribution can be used to estimate the quantum state~\cite{greenbaum_tomography}. To perform state tomography, we begin with a set of measurement operators $M$ and a density matrix $\rho$. Born's rule gives $\vec{p}_{i} = \text{Tr}(M^{\dag}_i M_i \rho)$. Where $p_i$ is the probability of obtaining a measurement result $M_i$ from the state $\rho$. This value can be determined experimentally by running the circuit and calculating the frequency with which $M_i$ is observed. The number of times the circuit is executed is termed the number of {\it shots}. A greater number of shots typically corresponds to a better estimate of $p_i$. By constructing a matrix $A = M^{\dag}_0 M_0 \otimes M^{\dag}_1 M_1 \hdots \otimes M^{\dag}_m M_m$, we can derive $A(\rho ^ {\otimes m}) = \vec{p}$. Inverting $A$ then reconstructs the density matrix $\rho$ given the measurement outcomes $\vec{p}$.

State tomography may be extended to determine the action of an operation on the system~\cite{merkel_tomography}. The Choi-Jamio\l kowski isomorphism may be used in open quantum systems to derive a map from states to processes~\cite{wood_open_quantum,Granade_2017}. This forms the basis of {\it quantum process tomography}. By selecting $2^{2n}$ linearly independent inputs $\rho_j$ we can characterise the action of a process $V$ on the system. The approach is much the same as state tomography, except we now perform the gate we are characterising prior to measurement $\vec{p}_{ij} = \text{Tr}(M^{\dag}_i M_i V\rho_j V^{\dag})$.

\subsection{Measurement Error Calibration} A more limited form of tomography can be used to characterise and invert measurement errors for devices with a fixed measurement basis (such as the IBM quantum devices)~\cite{Qiskit,nielsen}. For $n$ qubits, we construct a set of circuits that prepares and measures each of the $2^n$ possible states in the measurement basis. The columns of this {\it calibration matrix} $C$ (a subset of $\mathcal{E}$) contain the probabilities with which each measurement outcome was observed, while the rows correspond to the state that was prepared. This calibration matrix now describes the map from expected measurement states to observed measurement states, and hence describes the effect of measurement errors in the measurement basis. This calibration matrix can then be inverted and applied to mitigate noise on the device. This technique provides the most accurate characterisation of measurement errors, but comes with an exponential overhead in the number of calibration circuits.

If we assume that measurement errors are independent then our calibration matrix is linearly separable; $C_{1, 2, ... n} = C_{1} \otimes C_{2} \hdots \otimes C_{n}$ and can be constructed with only $2n$ calibration circuits. This method will characterise state-dependent measurement errors, however no information will be gained about multi-qubit correlated errors. Taking the assumptions of linear independence further, we can perform all of our calibrations using only two circuits; $I^{\otimes{n}}$ and $X^{\otimes{n}}$. The individual calibration matrices $C_{i}$ can be recovered from the joint calibration matrix by marginalising the contributions of the other qubits to the probability distribution; which takes the form of the normalised partial trace $|\text{Tr}_j(C_{ij})| \approx C_i$. This technique is limited as NISQ devices exhibit measurement crosstalk~\cite{jigsaw} and as a result measurement errors are not independent.

\subsection{Randomised Benchmarking (RB)} RB is a fundamentally different technique to tomography based calibration. A set of random circuits with the overall action $I$ of varying lengths are constructed. Each circuit is executed, and the probability of measuring $\ket{0}^{\otimes n}$ state dictates the average error rate of that circuit. The error rate is a function of the circuit depth, and by fitting error rates from random circuits of varying lengths we can estimate the average gate and SPAM errors on the device. Although it requires far fewer operations than tomographic methods, randomised benchmarking cannot distinguish correlated and state-dependent errors. This average error rate is useful for device profiling and some simulations, but is not as useful for implementing error mitigation strategies.

\textbf{Pauli / Clifford twirling}~\cite{harper_correlated} is an extension of Randomised benchmarking that approximates the error channel of a quantum device as a {\it Gibbs random field} (GRF)~\cite{harper_correlated}. It uses the difference between the fit to the GRF and the observed distribution to determine the incidence of correlated errors on the device. This method scales polynomially in the number of qubits $n$. Pauli twirling provides an efficient approximation of the SPAM free errors, but as a result does not include biased or correlated measurement errors.

\subsection{Circuit Specific Strategies}
Another class of approaches involves attempting to profile and minimise the noise at the whole circuit level. These methods depend on the choice of circuit and must be re-run for each new circuit, in contrast with schemes that directly characterise measurement errors on a device.

{\bf Static Invert and Measure} (SIM)~\cite{swamit_state_dep} targets state-dependent measurement errors on a particular circuit and tackles scalability by restricting itself to exactly four characterisation circuits. The circuit that is being characterised is executed, and prior to measurement one of the following four operators are applied: $I^{\otimes n}$, $X^{\otimes n}$, $(I\otimes X)^{\otimes \frac{n}{2}}$ and $(X \otimes I)^{\otimes \frac{n}{2}}$. The results of these circuits are then averaged to attempt to mitigate the action of any state-dependent error on the output. For a circuit with a state-dependent measurement bias, this method will reduce the error rate by approximately half as it averages between the dependent states. SIM may average over correlated measurement errors that align with its characterisation circuits.

{\bf Adaptive Invert and Measure} (AIM)~\cite{swamit_state_dep} follows a similar strategy to SIM in applying characterisation operators after the action of a target circuit. It increases the pool of characterisation circuits and attempts to determine a set of $k$ `correct' strings to use for averaging. It begins with $\frac{r_1 n}{2}$ characterisation circuits of the form $I^{\otimes 2i}\otimes X ^{\otimes 4} \otimes I^{\otimes n - 2i}$ for even $i$ in the range 0 to $\frac{n}{2}$ and applies them prior to measurement on a target circuit. These `patch' circuits are then effectively acting on sets of four qubits at a time with an overlap of two qubits between different patches. The outputs of these circuits are then averaged and the top $k$ characterisation circuits are selected and used in a further $r_2 k$ executions which are in turn averaged to produce the final output. This selection mechanism assumes that some elements of those top $k$ bit strings are improving the success probability of the circuit. 

{\bf JIGSAW}~\cite{jigsaw} is another circuit specific strategy. Unlike AIM and SIM, it targets correlated measurement errors by constructing Bayesian filters. Each filter is constructed by randomly dividing the set of measured qubits into `patches'. The measurement distribution of each patch then forms a sub-table which acts as a local Bayes filter on a global measurement distribution. These filters are applied by splitting the global measurement distribution and updating elements corresponding to the patch. This subset is then normalised and rejoined with the global distribution.

While JIGSAW is currently a state of the art method in scalable measurement error suppression, it suffers from inconsistent results: should one of the sub-tables contain a single value, the renormalisation of that subset will promote it to a state that is measured with probability 1. When convolved with the global measurement distribution, this can result in JIGSAW erroneously over-reporting states that occur with low probability. There are two avenues by which this can occur: (1) if a measurement outcome in the `global' table is sufficiently distant from all other outcomes, or (2) if not all results are measured in the subset circuit. The impact of this inconsistency then depends on the choice of subset circuits, the distribution of measurement results, the associated noise of the device and the order in which sub-tables are applied.

\section{Coupling Map Calibration (CMC)} 		
\label{section:Model}

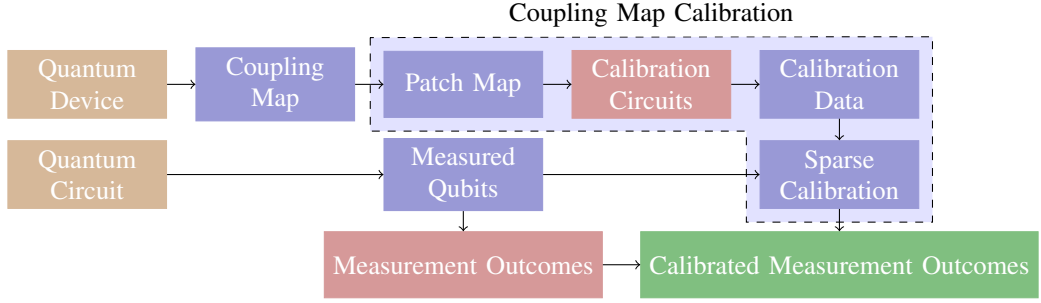
\begin{figure*}[t!]
%\centering
%\hspace{-1.9cm}
\centering
\begin{tikzpicture}
\node[rectangle,minimum height=2.5em, minimum width = 6em, yshift=-5mm, draw, orange!60!black!40, fill=orange!60!black!40, text=white, align=center] (device) {Quantum \\ Device};
\node[rectangle,minimum height=2.5em, minimum width = 6em, draw, right of = device, xshift=15mm, blue!60!black!40, fill=blue!60!black!40, text=white, align=center] (cmap) {Coupling \\ Map};
\draw[->] (device) to (cmap); 
\node[rectangle,minimum height=2.5em,  minimum width = 6em, draw, right of = cmap, xshift= 15mm, blue!60!black!40, fill=blue!60!black!40, text=white, align=center] (patches) {Patch Map};
\draw[->] (cmap) to (patches);
\node[rectangle,minimum height=2.5em,  minimum width = 6em, draw, right of = patches, xshift= 15mm, red!60!black!40, fill=red!60!black!40, text=white, align=center] (cdata) {Calibration \\ Circuits};
\draw[->] (patches) to (cdata);
\node[rectangle,minimum height=2.5em,  minimum width = 6em, draw, right of = cdata, xshift= 15mm, blue!60!black!40, fill=blue!60!black!40, text=white, align=center] (pairs) {Calibration \\ Data};
\draw[->] (cdata) to (pairs);

%\draw [decorate,decoration = {calligraphic brace}, minimum width = 3cm] (cmap.north west) --  (pairs.north east);

\node[rectangle,minimum height=2.5em, minimum width = 6em, draw, below of = device, yshift=-2mm, orange!60!black!40, fill=orange!60!black!40, text=white, align=center] (circ) {Quantum \\ Circuit};
\node[rectangle,minimum height=2.5em,  minimum width = 6em, draw, right of = circ, xshift=40mm, blue!60!black!40, fill=blue!60!black!40, text=white, align=center] (mqubits) {Measured \\ Qubits};
\draw[->] (circ) to (mqubits);
\node[rectangle,minimum height=2.5em,  minimum width = 6em, draw, right of = mqubits,  xshift=40mm, blue!60!black!40, fill=blue!60!black!40, text=white, align=center] (cpairs) {Sparse \\ Calibration};
\draw[->] (mqubits) to (cpairs);
\draw[->] (pairs) to (cpairs);

\node[rectangle,minimum height=2.5em, minimum width = 6em, draw, below of = mqubits, yshift=-2mm, red!60!black!40, fill=red!60!black!40, text=white, align=center] (circ_mdata) {Measurement Outcomes};
\draw[->] (mqubits) to (circ_mdata);
\node[rectangle,minimum height=2.5em, minimum width = 6em, draw, right of = circ_mdata, xshift=40mm, green!50!black!50, fill=green!50!black!50, text=white, align=center] (output) {Calibrated Measurement Outcomes};
\draw[->] (circ_mdata) to (output);
\draw[->] (cpairs) to (output);

\node[rectangle,minimum height=0em, minimum width = 6em, draw, above of = cdata, yshift=-0.8mm, white, fill=white, text=black, align=center] (CMC) {Coupling Map Calibration};

\begin{scope}[on background layer, local bounding box=T2] 
  \path[draw, dashed, fill = blue!15!white!75]
    ($(patches.north west)+(-0.5em,0.5em)$) 
    |- ($(patches.south west)+(0,-0.5em)$) 
    -| ($(cpairs.north west)+(-0.5em,-0.5em)$) 
    -| ($(cpairs.south west)+(-0.5em,-0.5em)$) 
    -| ($(cpairs.south east)+(0.5em,0.5em)$) 
    |- ($(cdata.north east)+(0.5em,0.5em)$) 
    -- cycle;
\end{scope}

\end{tikzpicture}
\caption{\small Coupling Map Calibration scheme for measurement error mitigation using patched calibrations. The scheme converts the coupling map of the device into a series of calibration patch circuits which are executed to collect data on two qubit calibrations. These calibrations can then be joined or traced over as required to mitigate measurement errors for any other circuit that is executed on the device. Red boxes indicate executions on a quantum device, blue elements are classical operations, orange inputs and green outputs. \label{fig:model}}
\end{figure*}

From the limitations of the current state of the art techniques discussed in the previous section, we derive three design objectives: 
\begin{enumerate}
    \item Characterising both correlated and state-dependent measurement errors on NISQ devices;
    \item Maintaining a polynomial number of characterisation shots in the number of qubits; and
    \item Demonstrating an improvement in the error rate reduction compared to the other polynomial methods, ideally achieving similar rates of error mitigation as calibration techniques.
\end{enumerate}
To achieve this, we propose a novel coupling map calibration (CMC) scheme, as outlined in Fig.~\ref{fig:model}. Previous work~\cite{harper_correlated} has established that the majority of correlated errors on modern quantum devices occur within physical proximity on the device. Leverage this insight we construct a set of sparse calibration matrices for each edge on the coupling map with each of these calibration edges termed a `patch'. By joining these sparse matrices we can then reconstruct an approximate global calibration matrix, which can then be used for measurement error mitigation. While the number of measurement circuits required to construct a full calibration matrix scales exponentially with the number of qubits, CMC only requires four such circuits for each edge and hence scales linearly in the number of edges on the coupling map. By the same argument we can perform non-local calibration circuits simultaneously and trace out the individual results. The number of calibration circuits is then reduced as a function of the sparsity of the coupling map.

\subsection{Coupling Map Calibration Patches}
Under the same assertion that correlated errors are physically correlated~\cite{harper_correlated}, physically distant qubits should not exhibit correlated errors. In other words, $C_{ij} = C_{i} \otimes C_{j}$ if $i$ and $j$ are sufficiently distant qubits. By the same argument, for pairs of adjacent qubits $ab$ and $cd$ that are sufficiently distant $C_{abcd} = C_{ab} \otimes C_{cd}$ if $ab$ and $cd$ are sufficiently distant pairs of qubits. From this, $C_{ab} = |\text{Tr}_{cd}(C_{abcd})|$ and $C_{cd} = |\text{Tr}_{ab}(C_{abcd})|$. If $ab$ and $cd$ are sufficiently distant edges on the coupling map, then we can calculate $C_{ab}$ and $C_{cd}$ simultaneously.

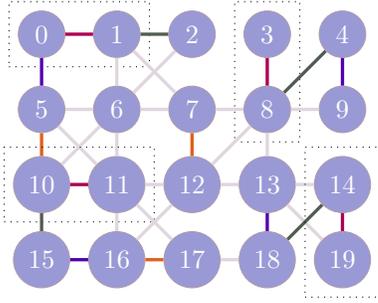
\begin{figure}
%\begin{subfigure}[b]{0.22\textwidth}
\centering
\begin{tikzpicture}
\node[draw, circle, violet!40!black!40, fill=blue!60!black!40, text=white] at (0, 5) (n_0) {$0$};
\node[draw, circle, violet!40!black!40, fill=blue!60!black!40, text=white] at (1, 5) (n_1) {$1$};
\node[draw, circle, violet!40!black!40, fill=blue!60!black!40, text=white] at (2, 5) (n_2) {$2$};
\node[draw, circle, violet!40!black!40, fill=blue!60!black!40, text=white] at (3, 5) (n_3) {$3$};
\node[draw, circle, violet!40!black!40, fill=blue!60!black!40, text=white] at (4, 5) (n_4) {$4$};
\node[draw, circle, violet!40!black!40, fill=blue!60!black!40, text=white] at (0, 4) (n_5) {$5$};
\node[draw, circle, violet!40!black!40, fill=blue!60!black!40, text=white] at (1, 4) (n_6) {$6$};
\node[draw, circle, violet!40!black!40, fill=blue!60!black!40, text=white] at (2, 4) (n_7) {$7$};
\node[draw, circle, violet!40!black!40, fill=blue!60!black!40, text=white] at (3, 4) (n_8) {$8$};
\node[draw, circle, violet!40!black!40, fill=blue!60!black!40, text=white] at (4, 4) (n_9) {$9$};
\node[draw, circle, violet!40!black!40, fill=blue!60!black!40, text=white] at (0, 3) (n_10) {$10$};
\node[draw, circle, violet!40!black!40, fill=blue!60!black!40, text=white] at (1, 3) (n_11) {$11$};
\node[draw, circle, violet!40!black!40, fill=blue!60!black!40, text=white] at (2, 3) (n_12) {$12$};
\node[draw, circle, violet!40!black!40, fill=blue!60!black!40, text=white] at (3, 3) (n_13) {$13$};
\node[draw, circle, violet!40!black!40, fill=blue!60!black!40, text=white] at (4, 3) (n_14) {$14$};
\node[draw, circle, violet!40!black!40, fill=blue!60!black!40, text=white] at (0, 2) (n_15) {$15$};
\node[draw, circle, violet!40!black!40, fill=blue!60!black!40, text=white] at (1, 2) (n_16) {$16$};
\node[draw, circle, violet!40!black!40, fill=blue!60!black!40, text=white] at (2, 2) (n_17) {$17$};
\node[draw, circle, violet!40!black!40, fill=blue!60!black!40, text=white] at (3, 2) (n_18) {$18$};
\node[draw, circle, violet!40!black!40, fill=blue!60!black!40, text=white] at (4, 2) (n_19) {$19$};

\draw [violet!64!red!100, line width=0.44mm,-] (n_0) -- (n_1);
\node [fit=(n_0) (n_1), draw, dotted] {};
\draw [violet!64!blue!100, line width=0.44mm,-] (n_0) -- (n_5);
%\node [fit=(n_0) (n_5), draw, dotted] {};
\draw [violet!64!green!100, line width=0.44mm,-] (n_1) -- (n_2);
\draw [violet!64!pink!54!black!19, line width=0.44mm,-] (n_1) -- (n_6);
\draw [violet!64!pink!54!black!19, line width=0.44mm,-] (n_1) -- (n_7);
\draw [violet!64!pink!54!black!19, line width=0.44mm,-] (n_2) -- (n_6);
\draw [violet!64!red!100, line width=0.44mm,-] (n_3) -- (n_8);
\node [fit=(n_3) (n_8), draw, dotted] {};
\draw [violet!64!green!100, line width=0.44mm,-] (n_4) -- (n_8);
\draw [violet!64!blue!100, line width=0.44mm,-] (n_4) -- (n_9);
\draw [violet!64!pink!54!black!19, line width=0.44mm,-] (n_5) -- (n_6);
\draw [violet!25!orange!100, line width=0.44mm,-] (n_5) -- (n_10);
\draw [violet!64!pink!54!black!19, line width=0.44mm,-] (n_5) -- (n_11);
\draw [violet!64!pink!54!black!19, line width=0.44mm,-] (n_6) -- (n_7);
\draw [violet!64!pink!54!black!19, line width=0.44mm,-] (n_6) -- (n_10);
\draw [violet!64!pink!54!black!19, line width=0.44mm,-] (n_6) -- (n_11);
\draw [violet!64!pink!54!black!19, line width=0.44mm,-] (n_7) -- (n_8);
\draw [violet!25!orange!100, line width=0.44mm,-] (n_7) -- (n_12);
\draw [violet!64!pink!54!black!19, line width=0.44mm,-] (n_8) -- (n_9);
\draw [violet!64!pink!54!black!19, line width=0.44mm,-] (n_8) -- (n_12);
\draw [violet!64!pink!54!black!19, line width=0.44mm,-] (n_8) -- (n_13);
\draw [violet!64!red!100, line width=0.44mm,-] (n_10) -- (n_11);
\node [fit=(n_10) (n_11), draw, dotted] {};
\draw [violet!64!green!100, line width=0.44mm,-] (n_10) -- (n_15);
\draw [violet!64!pink!54!black!19, line width=0.44mm,-] (n_11) -- (n_12);
\draw [violet!64!pink!54!black!19, line width=0.44mm,-] (n_11) -- (n_16);
\draw [violet!64!pink!54!black!19, line width=0.44mm,-] (n_11) -- (n_17);
\draw [violet!64!pink!54!black!19, line width=0.44mm,-] (n_12) -- (n_13);
\draw [violet!64!pink!54!black!19, line width=0.44mm,-] (n_12) -- (n_16);
\draw [violet!64!pink!54!black!19, line width=0.44mm,-] (n_13) -- (n_14);
\draw [violet!64!blue!100, line width=0.44mm,-] (n_13) -- (n_18);
\draw [violet!64!pink!54!black!19, line width=0.44mm,-] (n_13) -- (n_19);
\draw [violet!64!green!100, line width=0.44mm,-] (n_14) -- (n_18);
\draw [violet!64!red!100, line width=0.44mm,-] (n_14) -- (n_19);
\node [fit=(n_14) (n_19), draw, dotted] {};
\draw [violet!64!blue!100, line width=0.44mm,-] (n_15) -- (n_16);
\draw [violet!25!orange!100, line width=0.44mm,-] (n_16) -- (n_17);
\draw [violet!64!pink!54!black!19, line width=0.44mm,-] (n_17) -- (n_18);

\end{tikzpicture}
\caption{ \small Example measurement patches on the IBM Tokyo device with two qubits per patch and a distance between patches of at least one qubit and where each patch is represented by a colour. This pattern could be continued until all edges are incorporated in at least one patch.\label{fig:coupling_tokyo}}
\end{figure}

For this construction, we assume that all correlations are local on the device up to some distance $k$. Given the connectivity graph of the device, we can construct a set of `calibration patches' containing $n$ qubits such that all edges on the graph are included in at least one patch. We can then calibrate these two patches simultaneously without an increase in the number of shots (i.e., number of circuits executed on the quantum devices). 

From this strategy, the total number of shots per calibration matrix is $4r$, and the total number of calibration matrices is approximately a constant fraction of the number of edges in the coupling map, which in turn is typically on the order of the number of qubits on the device. For example, in the case of the IBM Tokyo device, the number of edges is $3$-$4$ times the number of qubits. A $k=1$ patching strategy reduces the number of circuits by approximately the same factor. This can be seen in Fig.~\ref{fig:coupling_tokyo}. We present a greedy construction for these patches in Algorithm~\ref{alg:coupling_map}. This approach iteratively finds a set of patches separated by at least distance $k$ until all edges of the graph are contained in a set. Each patch within a set may then be constructed simultaneously. The total cost in the number of measurements required is then only four times the number of such sets. For the Tokyo device, this corresponds to $40$ calibration circuits for all qubits individually, $140$ calibration circuits to characterise each edge individually, $54$ circuits for coupling map patching, $760$ circuits for all pairs of qubits, and $2^{20}$ circuits for the full calibration.

When tested on large random coupling maps ($>100$ qubits) with an average of four edges per qubit, this greedy approach reduced the number of calibration circuits by between a factor of 3 to 10. This style of coupling map (albeit not randomly constructed) is representative of the coupling maps of current IBMQ, Google, Rigetti, IonQ, and Honeywell devices~\cite{sycamore,rigetti,honeywell,Qiskit}.

\begin{algorithm}[t!]\small
    \caption{Greedy Distance $k$ Patch Construction}\label{alg:coupling_map}
    \begin{algorithmic}
        \STATE  $\mathrm{patch\_construct} (E, k)$
        \STATE Copy $E$ to $E'$
        \STATE Initialise an empty list $C$
        \WHILE{$E'$ is not empty}
            \STATE Initialise empty lists $C_i$, $B$
            \STATE Pop $e$ from $E'$ and append it to $C_i$ and $B$
            \STATE Mark all elements of $E$ as unvisited
            \WHILE {Not all vertices in $E$ are visited AND $B$ is not empty}
                \STATE Mark all elements of $B$ as visited
                \STATE Perform a depth $k$ BFS from each element in $B$, mark all vertices of up to distance $k$ as `visited'
                \STATE Replace $B$ with the depth $k$ boundary of the BFS
                \FOR{each edge $b$ in $B$}
                    \IF{If $b$ is not in $E'$ AND $b$ has not yet been visited}
                    \STATE Add $b$ to $C_i$
                    \STATE Mark $b$ as visited
                    \STATE Perform a depth $k$ BFS from each element in $B$, mark all vertices of up to distance $k$ as `visited'
                    \ENDIF 
                \ENDFOR
            \ENDWHILE
            \STATE Append $C_i$ to $C$
        \ENDWHILE
        \RETURN $C$
    \end{algorithmic}
\end{algorithm}

\subsection{Joining Calibration Matrices}
\label{sec:joining_calib}
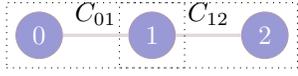
\begin{figure}
%\begin{subfigure}[b]{0.22\textwidth}
\centering
\begin{tikzpicture}
\node[draw, circle, violet!40!black!40, fill=blue!60!black!40, text=white] at (0, 0) (n_0) {$0$};
\node[draw, circle, violet!40!black!40, fill=blue!60!black!40, text=white] at (1.5, 0) (n_1) {$1$};
\node[draw, circle, violet!40!black!40, fill=blue!60!black!40, text=white] at (3, 0) (n_2) {$2$};

\draw [violet!64!pink!54!black!19, line width=0.44mm,-] (n_0) -- (n_1) node[pos=0.5, above, black] {$C_{01}$};
\draw [violet!64!pink!54!black!19, line width=0.44mm,-] (n_1) -- (n_2) node[pos=0.5, above, black] {$C_{12}$};

\node [fit=(n_0) (n_1), draw, dotted] {};
\node [fit=(n_1) (n_2), draw, dotted] {};
\end{tikzpicture}
\caption{Overlapping patches of two qubit calibrations; the goal of the method is to join these calibrations to construct an approximate $C_{012}$.\label{fig:patches}}
\end{figure}

The core of our model is finding a joint approximation of two overlapping local calibrations. Two disjoint calibrations may be joined via the tensor product: 
\begin{equation}
C_{ij} = C_i \otimes C_j
\end{equation}
However, this necessarily cannot capture any information regarding correlated errors between those two qubits. In order to capture correlated errors we must construct calibration circuits from the combination of all measurement operators across both qubits. Expanding this problem to $n$ qubits on a device scales as $2^n$. Ideally we would wish to perform two qubit calibrations with four calibration circuits each and stitch together these individual `patches' as seen in Fig.~\ref{fig:patches}. This raises the problem of how to join two-qubit calibration matrices with overlapping qubits. 
As calibration matrices are normalised maps between probability distributions, we can split a calibration matrix formed by the outer product of two other calibration matrices using the partial trace operation and normalising such that the sum of each column in the resulting calibration matrix is 1 as shown by
\begin{equation}
C_{i} = |\text{Tr}_j(C_i \otimes C_j)|.
\end{equation}
For a calibration matrix acting on multiple qubits, we can construct an approximate single-qubit calibration matrix using the same method
\begin{equation}
C_{i} \approx |\text{Tr}_j(C_{ij})|.
\end{equation}
If we have multiple multi-qubit calibration matrices all acting on the same qubit then we need to construct a technique by which we can apply each calibration jointly. For $v$ calibration matrices overlapping on some qubit labelled $j$, we can construct an approximation of the calibration matrix between qubits $i$ and $j$ (with some order parameter $v_a$) by
\begin{equation}
    C'_{ij}(v_a) = \left(I \otimes C_j^{\frac{v - 1 - v_a}{v}}\right)^{-1} C_{ij} \left(I \otimes C_j^{\frac{v_a}{v}}\right)^{-1}
\end{equation}
such that $\text{Tr}_i(C'_{ij}) \approx C_j^{\frac{1}{v}}$ and $\text{Tr}_{j}(C'_{ij}) \approx \text{Tr}_{j}(C_{ij})$. This construction can be seen in Fig~\ref{fig:cij_construct}. 

\begin{figure}
\centering
\begin{tikzpicture}
\node[rectangle,minimum height=2em,draw, blue!60!black!40, fill=blue!60!black!40, text=white] (cn) {$C_{j}^{1/v}$};
\node[above of=cn, yshift=-5mm] (tn) {$\otimes$};
\node[above of=tn, yshift=-5mm] (in) {$I$};

\node[right of=tn, xshift=-1mm] (in1) {$\hdots$};

\node[rectangle,minimum height=2em,draw, blue!60!black!40, fill=blue!60!black!40, text=white, right of=cn, xshift=9mm] (cn2) {$C_{j}^{1/v}$};
\node[above of=cn2, yshift=-5mm] (tn2) {$\otimes$};
\node[above of=tn2, yshift=-5mm] (in2) {$I$};

\node[rectangle,minimum height=5em,draw, blue!60!black!40, fill=blue!60!black!40, text=white, right of=tn2, xshift=2mm] (cijv) {$C'_{ij}(v_a)$};

\node[right of=cijv, xshift=1mm] (tn3) {$\otimes$};
\node[rectangle,minimum height=2em, yshift=5mm, draw, blue!60!black!40, fill=blue!60!black!40, text=white, below of=tn3, xshift=1mm] (cn3) {$C_{j}^{1/v}$};
\node[above of=tn3, yshift=-5mm] (in3) {$I$};

\node[right of=tn3] (in4) {$\hdots$};

\node[right of=in4, xshift=-2mm] (tn5) {$\otimes$};
\node[rectangle,minimum height=2em, yshift=5mm, draw, blue!60!black!40, fill=blue!60!black!40, text=white, below of=tn5, xshift=1mm] (cn5) {$C_{j}^{1/v}$};
\node[above of=tn5, yshift=-5mm] (in5) {$I$};

\node[left of=tn, xshift=1mm] (eq) {$:=$};

\node[rectangle,minimum height=5em,draw, blue!60!black!40, fill=blue!60!black!40, text=white, left of=eq, xshift=1mm] (cij) {$C_{ij}$};

\draw [decorate,decoration = {calligraphic brace}, minimum width = 3cm] (in.north west) --  (in2.north east);
\draw [decorate,decoration = {calligraphic brace}, minimum width = 3cm] (in3.north west) --  (in5.north east);

\node[above of = in1, black, yshift=2mm] {$v - v_a - 1$};
\node[above of = in4, black, yshift=2mm, xshift=-1mm] {$v_a$};

\end{tikzpicture}
\caption{\small Construction of $C'{ij}(v_a)$ given $C_{ij}$ and a choice of order parameter $v_a$ and a total number of calibration matrices to join over this index such that $|\text{Tr}_{i}(C'_{ij})| \approx C_{j}^{\frac{1}{n}}$, and $\text{Tr}_{j}(C'_{ij}) \approx \text{Tr}_{j}(C_{ij})$. When building the overall calibration matrix other pairwise calibrations acting on $j$ with their own order parameters $v_{0\hdots v - 1}$ will fill out the other $v - 1$ terms.\label{fig:cij_construct}}
\end{figure}

Similarly, for some other qubit $k$ with order parameter $v_b$ we can construct
\begin{equation}
    C'_{jk}(v_b) = \left( C_j^{\frac{v - 1 - v_b}{v}} \otimes I \right)^{-1} C_{jk} \left(C_j^{\frac{v_b}{v}} \otimes I\right)^{-1}
\end{equation}
such that if $v_1 > v_0$ {and $v = 2$
\begin{equation}
    C'_{ijk} = (I \otimes C'_{jk})(C'_{ij} \otimes I).
\end{equation}
This allows us to patch together and construct a joint calibration matrix from the overlapping individual calibrations given an explicit order.

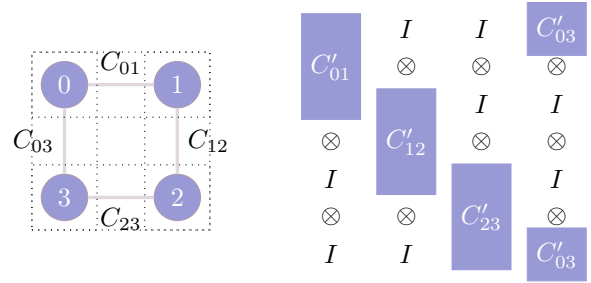
\begin{figure}
\begin{subfigure}[b]{0.22\textwidth}
\centering
\begin{tikzpicture}
\node[draw, circle, violet!40!black!40, fill=blue!60!black!40, text=white] at (0, 0) (n_0) {$0$};
\node[draw, circle, violet!40!black!40, fill=blue!60!black!40, text=white] at (1.5, 0) (n_1) {$1$};
\node[draw, circle, violet!40!black!40, fill=blue!60!black!40, text=white] at (1.5, -1.5) (n_2) {$2$};
\node[draw, circle, violet!40!black!40, fill=blue!60!black!40, text=white] at (0, -1.5) (n_3) {$3$};

\node[] at (0, -2.5) (n_space) {};

\draw [violet!64!pink!54!black!19, line width=0.44mm,-] (n_0) -- (n_1) node[pos=0.5, above, black] {$C_{01}$};
\draw [violet!64!pink!54!black!19, line width=0.44mm,-] (n_1) -- (n_2) node[pos=0.5, right, black] {$C_{12}$};
\draw [violet!64!pink!54!black!19, line width=0.44mm,-] (n_2) -- (n_3) node[pos=0.5, below, black] {$C_{23}$};
\draw [violet!64!pink!54!black!19, line width=0.44mm,-] (n_3) -- (n_0) node[pos=0.5, left, black] {$C_{03}$};

\node [fit=(n_0) (n_1), draw, dotted] {};
\node [fit=(n_1) (n_2), draw, dotted] {};
\node [fit=(n_2) (n_3), draw, dotted] {};
\node [fit=(n_3) (n_0), draw, dotted] {};
\end{tikzpicture}
\end{subfigure}
\begin{subfigure}[b]{0.24\textwidth}
\centering
\begin{tikzpicture}
\node[rectangle,minimum height=4em,draw, blue!60!black!40, fill=blue!60!black!40, text=white] (c01) at (0, 0) {$C'_{01}$};
\node[below of=c01] (t02) {$\otimes$};
\node[below of=t02, yshift=5mm] (i02) {$I$};
\node[below of=i02, yshift=5mm] (t03) {$\otimes$};
\node[below of=t03, yshift=5mm] (i03) {$I$};

\node[rectangle,minimum height=4em,draw, right of=t02, blue!60!black!40, fill=blue!60!black!40, text=white] (c12) {$C'_{12}$};

\node[above of=c12] (t10) {$\otimes$};
\node[above of=t10, yshift=-5mm] (i02) {$I$};

\node[below of=c12] (t13) {$\otimes$};
\node[below of=t13, yshift=5mm] (i13) {$I$};

\node[rectangle,minimum height=4em,draw, right of=t13, blue!60!black!40, fill=blue!60!black!40, text=white] (c23) {$C'_{23}$};

\node[above of=c23] (t21) {$\otimes$};
\node[above of=t21, yshift=-5mm] (i21) {$I$};
\node[above of=i21, yshift=-5mm] (t20) {$\otimes$};
\node[above of=t20, yshift=-5mm] (i20) {$I$};

\node[rectangle,minimum height=2em,draw, right of=i20, blue!60!black!40, fill=blue!60!black!40, text=white] (c30) {$C'_{03}$};

\node[below of=c30, yshift=5mm] (t31) {$\otimes$};
\node[below of=t31, yshift=5mm] (i31) {$I$};
\node[below of=i31, yshift=5mm] (t32) {$\otimes$};
\node[below of=t32, yshift=5mm] (i32) {$I$};
\node[below of=i32, yshift=5mm] (t33) {$\otimes$};

\node[rectangle,minimum height=2em,draw, below of=t33, yshift=5mm, blue!60!black!40, fill=blue!60!black!40, text=white] (c33) {$C'_{03}$};

\end{tikzpicture}
\end{subfigure}
\caption{\small A spanning set of two qubit patches incorporating all edges in the graph and the form of the associated calibration matrix. Columns represent tensor product while rows represent matrix products. Each individual column is itself a sparse matrix. For large systems (with matrices scaling with the number of qubits as $2^n \times 2^n$) it is faster to do a series of sparse matrix vector multiplications than it is to perform a single dense matrix multiplication.\label{fig:spanning_set}}
\end{figure}

This method can be extended to joining calibration matrices of arbitrary sizes and with overlaps by selecting the appropriate order parameter for each contributing term in $C'$. An example of this for a square plaquette of connected qubits can be seen in Fig.~\ref{fig:spanning_set}.

Armed with the ability to approximate larger calibration matrices from smaller ones, we can now build a calibration matrix for the system. To construct this, we need a spanning set of calibration matrices of $n$ qubits across the system, before using the method above to construct a set of calibrations that may be applied linearly. 

As correlated errors tend to be spatially close on the coupling map~\cite{harper_correlated}, the question then becomes one of deciding which spanning set of calibration matrices should be used. For a base application of CMC we have opted to use edges of the coupling map as the spanning set of calibration matrices.

\subsection{CMC: Mitigating Measurement Errors}

Using this ability to construct and join sets of calibration matrices we may now consider how to wrangle this into a measurement error mitigation scheme. We start with the coupling map of the device and the set of qubits that we are measuring. If the measured qubit $i$ has no neighbours, then we can construct the effective calibration matrix from the patches $C_{ij}$ by tracing out the non-measured qubits $j$ such that $C'_i = |\prod_j \text{Tr}_j(C_{ij})^{\frac{1}{v}}|$. If instead qubit $i$ shares a non-zero number of edges we follow the method discussed in section~\ref{sec:joining_calib} and trace out over edges shared with non-measured qubits while multiplying the shared edges. In both cases, the order in which these matrices are applied must be identical to the order in which the calibration matrices were constructed as seen in Fig.~\ref{fig:cij_construct}.

Once this set of sparse calibration matrices have been constructed, we can then invert each matrix then take the tensor product with the identity for each non-participating qubit. By reversing the order of these inverted matrices we have constructed the inverse of the calibration matrix. This set of inverted calibration matrices can now be applied to a measurement distribution in order to perform the desired measurement error mitigation. We term this scheme CMC.

If the number of qubits in a calibration patch is significantly less than the number of qubits in the system, then each individual calibration matrix will be sparse. The maximum number of entries in the measurement vector is bounded by the number of shots performed on the system, and can be periodically culled of very low weight entries. In the regime of a $50$+ qubit system, applying a series of sparse matrix-vector products is much more performant than a $2^{n}\times2^{n}$ dense full calibration matrix.

\subsection{ERR: Device Tailored Mitigations}

As seen in Fig.~\ref{fig:corr_coupling_demo}, while highly correlated errors tend to be local, they do not necessarily align to the coupling map. If a device exhibits a systemic correlated error that persists through multiple calibrations, we may modify our coupling map method to instead characterise the noisiest edges. The goal is to construct a coupling map with at most $n$ edges while attempting to maximise the number of highly correlated measurement errors. We present {\it ERR} an $O(n^2)$ greedy approach to this problem in Algorithm~\ref{alg:err_coupling_map} with a graphical example in Fig.~\ref{fig:err_map_demo}. Here $E$ is the initial coupling map, and $k$ is a locality parameter, such that only two-qubit edges of distance less than $k$ are considered.
\begin{algorithm}[t!]\small
    \caption{Error Coupling Map Construction}\label{alg:err_coupling_map}
    \begin{algorithmic}
        \STATE  $\mathrm{error\_coupling\_map\_construct} (E, k)$
        \STATE Construct all 1 and 2 qubit calibration matrices $C_i$, $C_{i < j}, \forall |i, j| < k \in E$
        \STATE Calculate all edge weights $w_{i, j} = ||C_i \otimes C_j - C_{ij}||$
        \STATE Sort edge weights $W$ in descending order $w_{i, j}$
        \STATE Initialise an empty graph $E'$
        \FOR{$w_{i, j}$ in $W$ and $|E'| < n$}
            \IF{$i$ in $E'$ and $j$ not in $E'$}
            \STATE Add $j$ and edge $i, j$ to $E'$
            \ENDIF
            \IF{$j$ in $E'$ and $i$ not in $E'$}
            \STATE Add $i$ and edge $i, j$ to $E'$
            \ENDIF
            \IF{$i$ and $j$ not in $E'$}
            \STATE Let $w_i$, $w_j$ be the next weighted edges containing $i$ and $j$
            \STATE Add $\min_{w}((i, w_i), (j, w_j))$ and dangling edge $i, j$ to $E'$ 
            \ENDIF
        \ENDFOR
        \STATE return $E'$
    \end{algorithmic}
\end{algorithm}
Note that unlike the computational coupling map there is no requirement that this error coupling map be connected. We can then perform  CMC over this error coupling map to construct our sparse measurement calibration matrices.

\begin{figure}[t!]
\begin{subfigure}[b]{0.15\textwidth}
\centering
\begin{tikzpicture}
\def \radius {0.9cm}
\node[draw, circle, violet!40!pink!40!black!40, fill=blue!60!black!40, text=white] (n_0) {$0$};
\node[draw, circle, violet!40!pink!40!black!40, fill=blue!60!black!40, text=white, right of = n_0, yshift=-0.75em]  (n_1) {$1$};
\node[draw, circle, violet!40!pink!40!black!40, fill=blue!60!black!40, text=white, right of = n_1, yshift=0.75em]  (n_2) {$2$};
\node[draw, circle, violet!40!pink!40!black!40, fill=blue!60!black!40, text=white, below of = n_1, yshift=0.75em]  (n_3) {$3$};
\node[draw, circle, violet!40!pink!40!black!40, fill=blue!60!black!40, text=white, below of = n_3, yshift=0.75em]  (n_5) {$5$};
\node[draw, circle, violet!40!pink!40!black!40, fill=blue!60!black!40, text=white, left of = n_5, yshift = -0.75em]  (n_4) {$4$};
\node[draw, circle, violet!40!pink!40!black!40, fill=blue!60!black!40, text=white, right of = n_5, yshift = -0.75em]  (n_6) {$6$};

\draw [violet!55!black!45,line width=0.63mm,-] (n_0) -- (n_1);
\draw [violet!55!black!45,line width=0.63mm,-] (n_1) -- (n_2);
\draw [violet!55!black!45,line width=0.63mm,-] (n_1) -- (n_3);
\draw [violet!55!black!45,line width=0.63mm,-] (n_3) -- (n_5);
\draw [violet!55!black!45,line width=0.63mm,-] (n_5) -- (n_4);
\draw [violet!55!black!45,line width=0.63mm,-] (n_5) -- (n_6);
\end{tikzpicture}
\caption{\scriptsize Coupling Map}
\end{subfigure}
\hfill
\begin{subfigure}[b]{0.15\textwidth}
\centering
\begin{tikzpicture}
\def \radius {0.9cm}
\node[draw, circle, violet!40!pink!40!black!40, fill=blue!60!black!40, text=white] (n_0) {$0$};
\node[draw, circle, violet!40!pink!40!black!40, fill=blue!60!black!40, text=white, right of = n_0, yshift=-0.75em]  (n_1) {$1$};
\node[draw, circle, violet!40!pink!40!black!40, fill=blue!60!black!40, text=white, right of = n_1, yshift=0.75em]  (n_2) {$2$};
\node[draw, circle, violet!40!pink!40!black!40, fill=blue!60!black!40, text=white, below of = n_1, yshift=0.75em]  (n_3) {$3$};
\node[draw, circle, violet!40!pink!40!black!40, fill=blue!60!black!40, text=white, below of = n_3, yshift=0.75em]  (n_5) {$5$};
\node[draw, circle, violet!40!pink!40!black!40, fill=blue!60!black!40, text=white, left of = n_5, yshift = -0.75em]  (n_4) {$4$};
\node[draw, circle, violet!40!pink!40!black!40, fill=blue!60!black!40, text=white, right of = n_5, yshift = -0.75em]  (n_6) {$6$};

\draw [violet!55!black!45,line width=0.63mm,-] (n_0) -- (n_1);
\draw [violet!20!black!80,line width=1.377mm,-] (n_0) to [bend left=15] (n_2);
\draw [violet!48!black!52,line width=0.726mm,-] (n_0) -- (n_3);
%\draw [violet!70!black!30,line width=0.408mm,-] (n_0) to [bend left=15] (n_4);
\draw [violet!69!black!31,line width=0.435mm,-] (n_0) -- (n_5);
%\draw [violet!59!black!41,line width=0.573mm,-] (n_0) -- (n_6);
\draw [violet!55!black!45,line width=0.618mm,-] (n_1) -- (n_2);
\draw [violet!38!black!62,line width=0.864mm,-] (n_1) -- (n_3);
\draw [violet!64!black!36,line width=0.495mm,-] (n_1) -- (n_4);
\draw [violet!61!black!39,line width=0.543mm,-] (n_1) to [bend left=60] (n_5);
\draw [violet!61!black!39,line width=0.5399999999999999mm,-] (n_1) -- (n_6);
\draw [violet!41!black!59,line width=0.8220000000000001mm,-] (n_2) -- (n_3);
%\draw [violet!67!black!33,line width=0.462mm,-] (n_2) -- (n_4);
\draw [violet!66!black!34,line width=0.462mm,-] (n_2) -- (n_5);
%\draw [violet!69!black!31,line width=0.423mm,-] (n_2) to [bend right=15] (n_6);
\draw [violet!37!black!63,line width=0.873mm,-] (n_3) -- (n_4);
\draw [violet!20!black!80,line width=1.4040000000000001mm,-] (n_3) -- (n_5);
\draw [violet!33!black!67,line width=0.9299999999999999mm,-] (n_3) -- (n_6);
\draw [violet!67!black!33,line width=0.453mm,-] (n_4) -- (n_5);
\draw [violet!20!black!80,line width=1.401mm,-] (n_4) to [bend right=15] (n_6);
\draw [violet!61!black!39,line width=0.543mm,-] (n_5) -- (n_6);
\end{tikzpicture}
\caption{\scriptsize $k\le3$ Correlations}
\end{subfigure}
\hfill
\begin{subfigure}[b]{0.15\textwidth}
\centering
\begin{tikzpicture}
\def \radius {0.9cm}
\node[draw, circle, violet!40!pink!40!black!40, fill=blue!60!black!40, text=white] (n_0) {$0$};
\node[draw, circle, violet!40!pink!40!black!40, fill=blue!60!black!40, text=white, right of = n_0, yshift=-0.75em]  (n_1) {$1$};
\node[draw, circle, violet!40!pink!40!black!40, fill=blue!60!black!40, text=white, right of = n_1, yshift=0.75em]  (n_2) {$2$};
\node[draw, circle, violet!40!pink!40!black!40, fill=blue!60!black!40, text=white, below of = n_1, yshift=0.75em]  (n_3) {$3$};
\node[draw, circle, violet!40!pink!40!black!40, fill=blue!60!black!40, text=white, below of = n_3, yshift=0.75em]  (n_5) {$5$};
\node[draw, circle, violet!40!pink!40!black!40, fill=blue!60!black!40, text=white, left of = n_5, yshift = -0.75em]  (n_4) {$4$};
\node[draw, circle, violet!40!pink!40!black!40, fill=blue!60!black!40, text=white, right of = n_5, yshift = -0.75em]  (n_6) {$6$};

\draw [violet!20!black!80,line width=1.377mm,-] (n_0) to [bend left=15] (n_2);
\draw [violet!38!black!62,line width=0.864mm,-] (n_1) -- (n_3);
\draw [violet!41!black!59,line width=0.8220000000000001mm,-] (n_2) -- (n_3);
\draw [violet!37!black!63,line width=0.873mm,-] (n_3) -- (n_4);
\draw [violet!20!black!80,line width=1.4040000000000001mm,-] (n_3) -- (n_5);
\draw [violet!33!black!67,line width=0.9299999999999999mm,-] (n_3) -- (n_6);
\draw [violet!20!black!80,line width=1.401mm,-] (n_4) to [bend right=15] (n_6);
\end{tikzpicture}
\caption{\scriptsize Error Map}
\end{subfigure}
\caption{\small Construction of a $k=3$ local error map using ERR\label{fig:err_map_demo}.}
\end{figure}
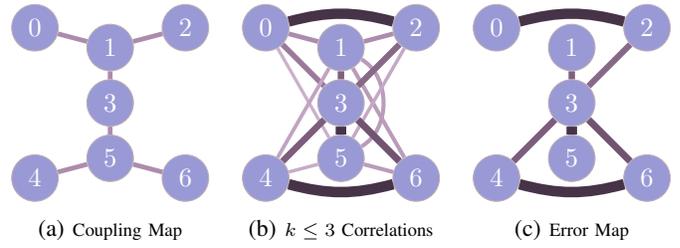

\section{Evaluation Methodology}
\label{section:Methodology}

\begin{figure}
\begin{subfigure}[b]{0.49\textwidth}
\begin{subfigure}[b]{0.24\textwidth}
\centering\hspace{0.45cm}
\includegraphics[width=1\linewidth]{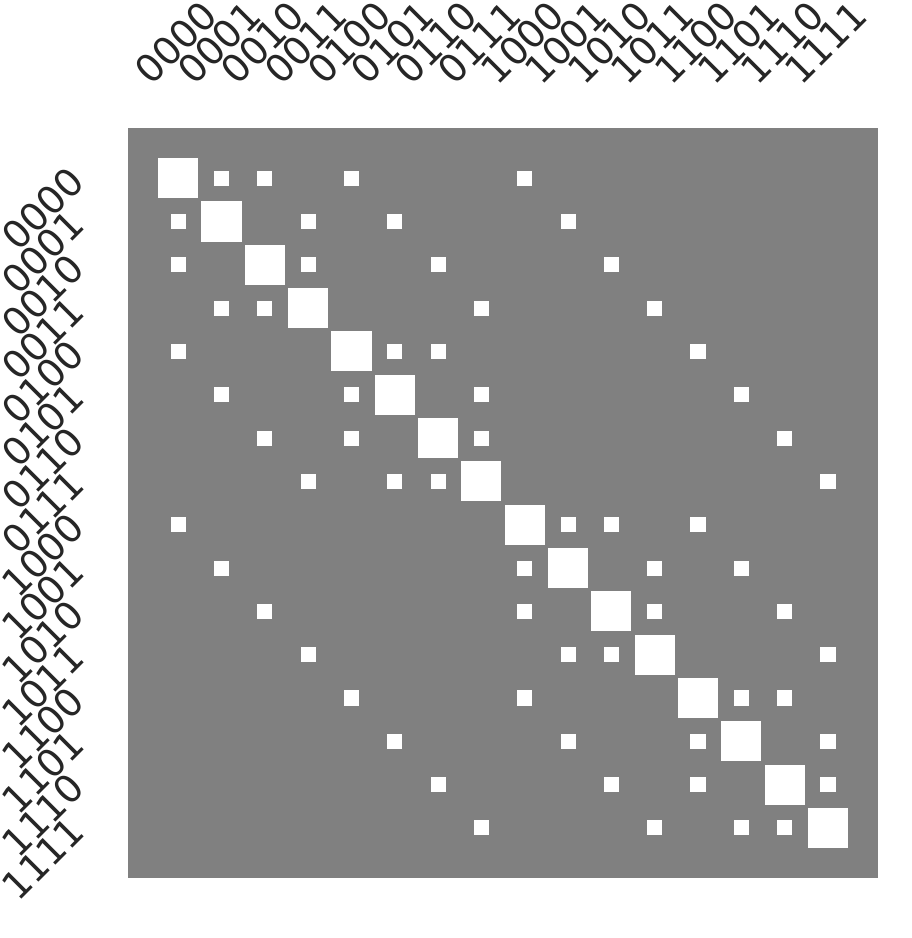}
%\caption{\small Distance 1 state-dependent Errors \label{subfig:biased_mit}}
\end{subfigure}
\begin{subfigure}[b]{0.24\textwidth}\hspace{-0.3cm}
\includegraphics[width=1\linewidth]{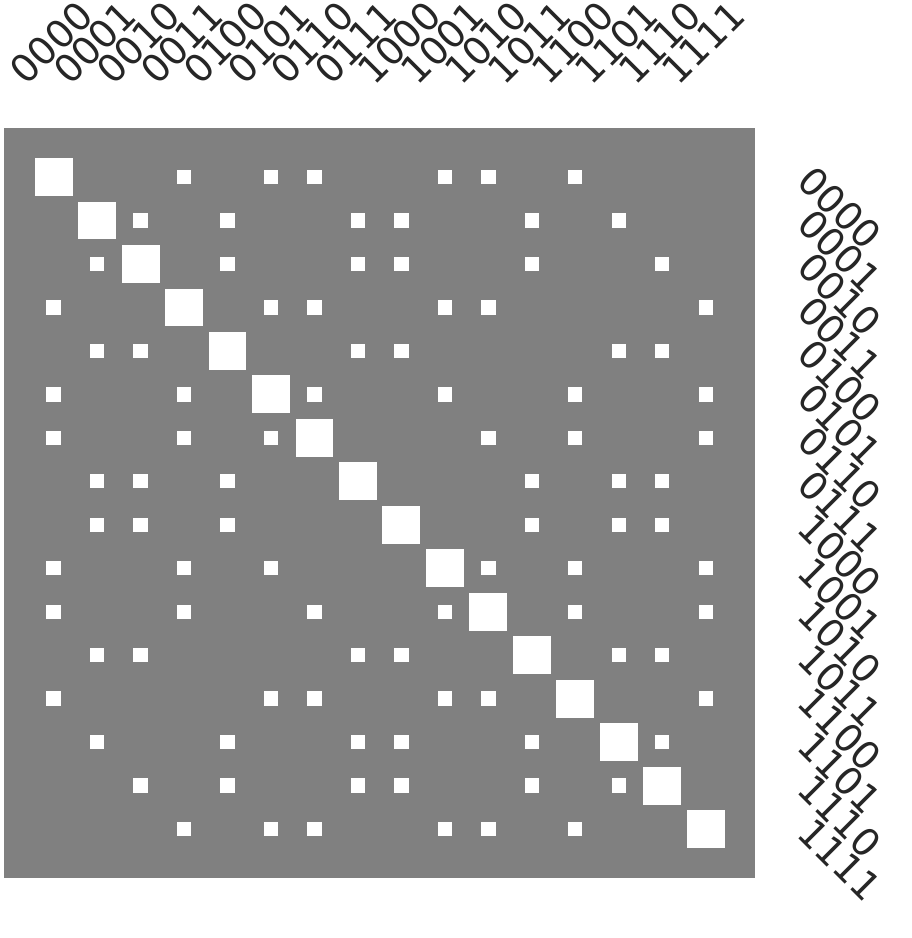}
%\caption{\small Distance 2 state-dependent Errors \label{subfig:biased_mit}}
\end{subfigure}
\begin{subfigure}[b]{0.24\textwidth}
\centering\hspace{0.45cm}
\includegraphics[width=1\linewidth]{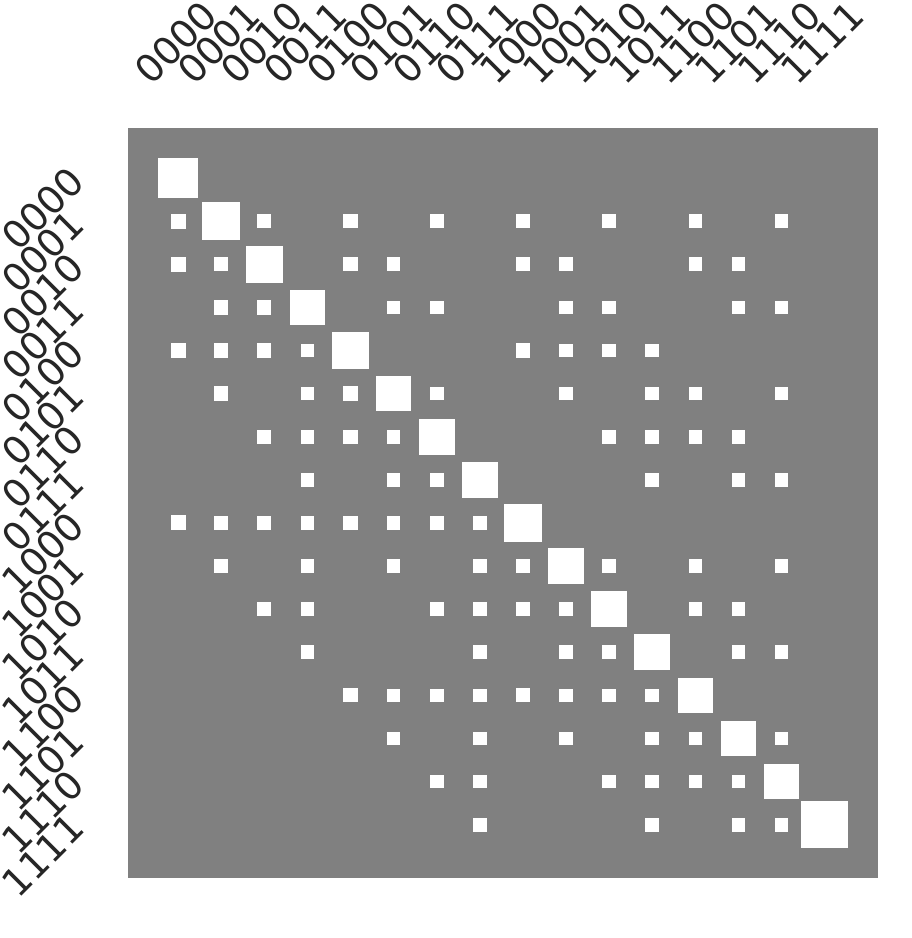}
%\caption{\small Distance 1 state-dependent Errors \label{subfig:biased_mit}}
\end{subfigure}
\begin{subfigure}[b]{0.24\textwidth}\hspace{-0.3cm}
\includegraphics[width=1\linewidth]{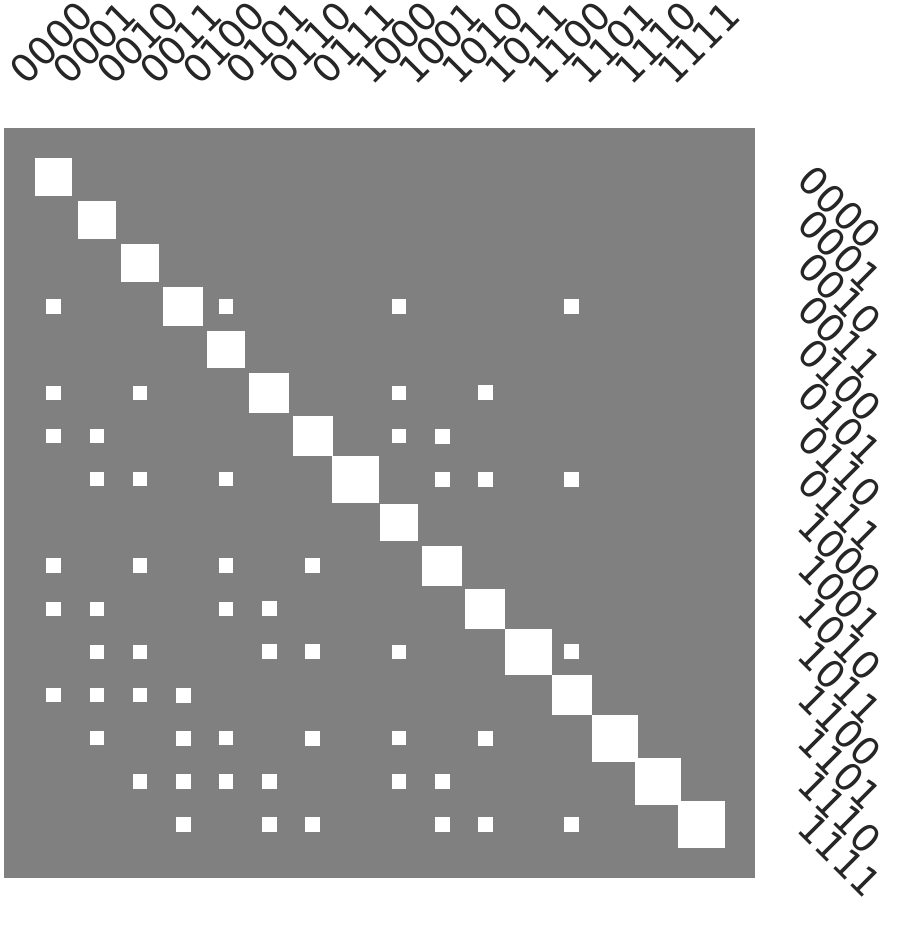}
%\caption{\small Distance 2 state-dependent Errors \label{subfig:biased_mit}}
\end{subfigure}
\end{subfigure}
\begin{subfigure}[b]{0.49\textwidth}
\begin{subfigure}[b]{0.24\textwidth}
\centering\hspace{0.45cm}
\includegraphics[width=1\linewidth]{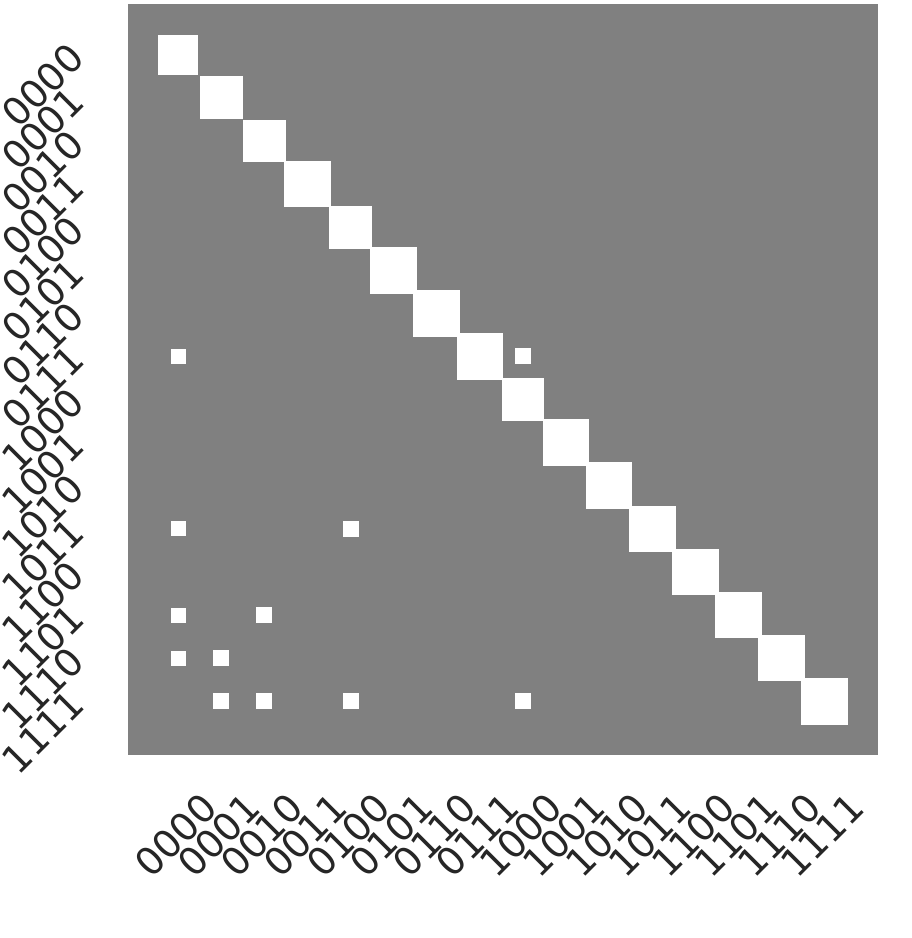}
%\caption{\small Distance 3 state-dependent Errors \label{subfig:biased_mit}}
\end{subfigure}
\begin{subfigure}[b]{0.24\textwidth}\hspace{-0.3cm}
\includegraphics[width=1\linewidth]{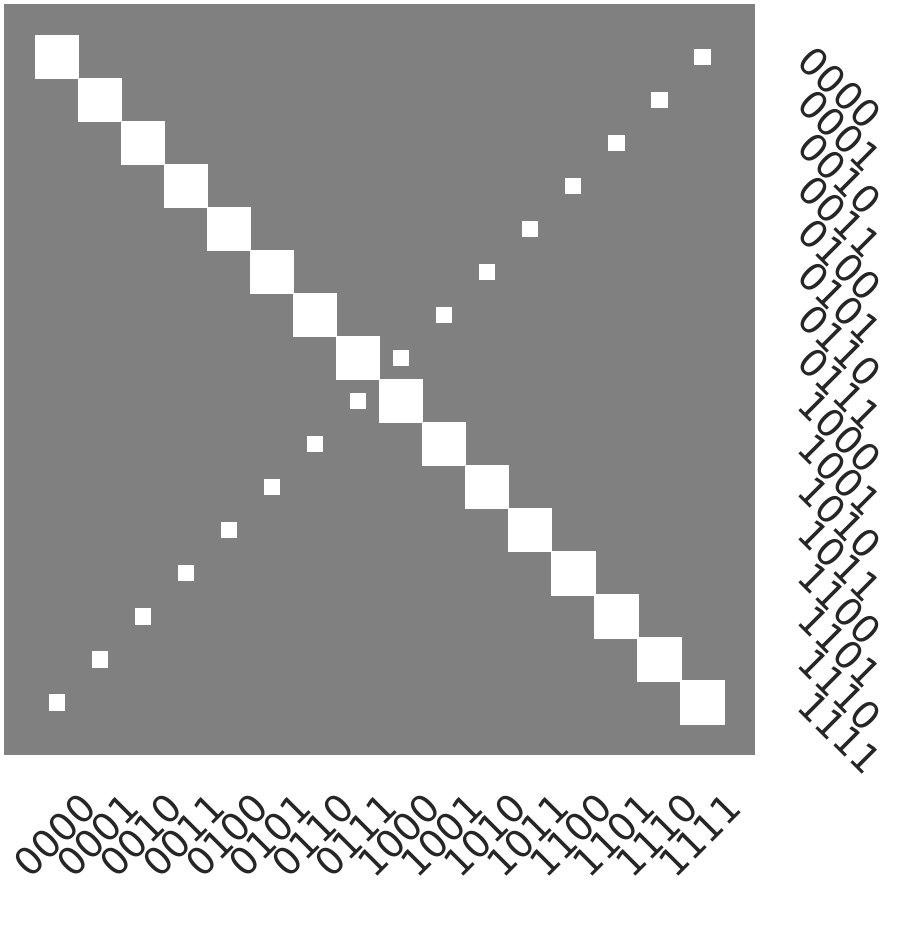}
%\caption{\small Distance 4 state-dependent Errors\label{subfig:biased_mit}}
\end{subfigure}
\begin{subfigure}[b]{0.24\textwidth}
\centering\hspace{0.45cm}
\includegraphics[width=1\linewidth]{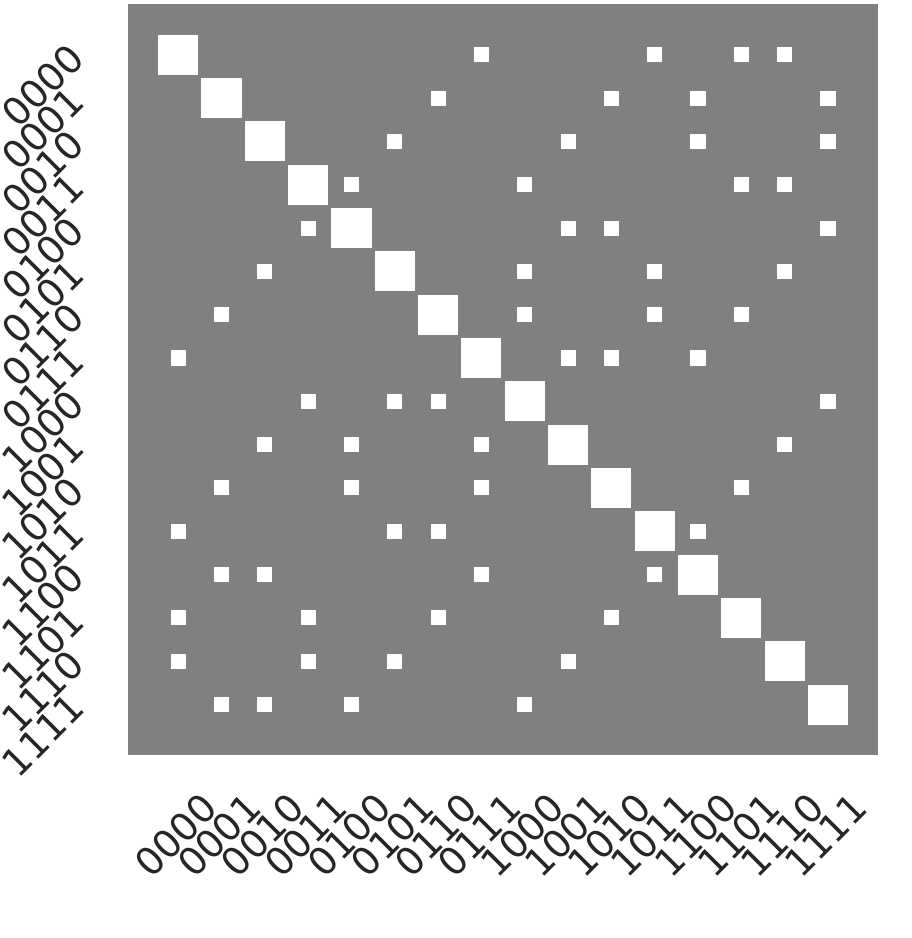}
%\caption{\small Distance 3 state-dependent Errors \label{subfig:biased_mit}}
\end{subfigure}
\begin{subfigure}[b]{0.24\textwidth}\hspace{-0.3cm}
\includegraphics[width=1\linewidth]{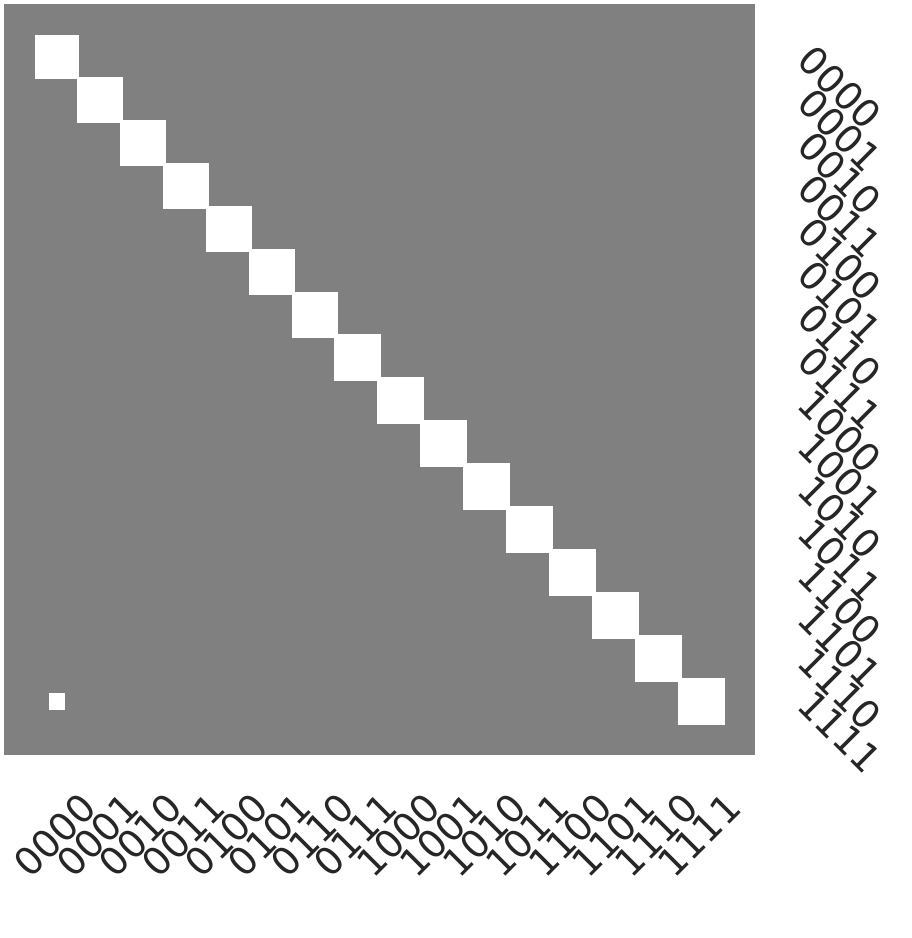}
%\caption{\small Distance 4 state-dependent Errors\label{subfig:biased_mit}}
\end{subfigure}
\end{subfigure}
\caption{\small (Left) Hinton diagrams of simulated correlated measurement errors over four qubits. Clockwise from top left; single qubit (uncorrelated), two qubit (all pairs), three qubit (triplets), four qubit (flip all bits). The four qubit error channel simply flips all the bits in the state. (Right) Hinton diagrams of simulated state-dependent measurement errors over four qubits. The four-qubit channel only has a single non-diagonal entry as there is only one four-qubit state-dependent measurement error.\label{fig:hinton_corr}}
\end{figure}

This section will discuss the application and methodology of the designs discussed in Section \ref{section:Model} and the construction of the circuits that are executed on the IBMQ devices. These circuits will be used to compare the following measurement error mitigation techniques: SIM, AIM, Full calibration, Linear calibration, CMC, CMC-ERR, and JIGSAW.

In this paper, we conducted experiments on the IBMQ, `Quito' `Lima', `Manila', and `Nairobi' devices. The relevant figures of merit are the overall error rate (here taken as the probability with which the correct output is produced) and the total number of quantum device executions. The goal is to minimise the error rate for a fixed number of measurement shots when comparing circuit selection and averaging techniques (AIM/SIM) against classical post-processing approaches (calibration methods). We define the success probability as the frequency with which the measurement output aligns with a classically verified error-free result. An extension of this is the one norm distance which measures the difference between a classically verified distribution of measurement outcomes and an observed measurement distribution.

To compare the various measurement error mitigation techniques, we focus on the GHZ benchmark. This benchmark is the smallest circuit that entangles all qubits on a device. 

\subsection{Benchmark Error Simulations}

The first benchmarks are a set of simulated measurement errors over a range of states. The simulated error channels exhibit varying amounts of state-dependent, uniform, and correlated errors. These simulated error distributions can be seen in Fig~\ref{fig:hinton_corr}. These diagrams indicate the probability with which a particular input state is mapped to an output state by the measurement error channel. The size of the squares scales with the probability of that operation. By constructing combinations of these error channels, we compare different mitigation methods against a range of scenarios relating purely to measurement errors. Using these measurement error channels, we can construct errors that probe the responsiveness of different error mitigation techniques to correlated and state-dependent errors. We can also determine the scalability of each of these methods in terms of the total number of shots required to produce a consistent result.

Simulations are then performed by applying the matrices associated with each gate and measurement operator to construct the ideal, noiseless, measurement output vector. We then apply the constructed measurement error channel to this output vector, producing a vector of the probabilities of measuring each state given the measurement error channel. This distribution can then be sampled to produce simulated measurement results.

In addition to these tailored measurement errors, we construct a set of simulated backend devices as seen in Fig.~\ref{fig:simulated_device_backends}. The topologies of these simulated devices emulate the coupling map architectures of a range of modern quantum devices. We use these architectures along with the Qiskit statevector simulator~\cite{Qiskit} to evaluate GHZ circuits for varying device sizes. For these simulations, we set a one qubit gate error rate of $0.1\%$, a two qubit gate error rate of $1\%$, and a readout error rate between $2-8\%$, and state-dependent measurement errors between $2-8\%$ for each qubit for both $\ket{0} \rightarrow \ket{1}$ and $\ket{0} \rightarrow \ket{1}$. $T_1$ and $T_2$ times are set to infinity. These single and two-qubit error rates are approximately analogous to those exhibited by the current NISQ devices; for example, one calibration of the IBM Quito device reports an average of two-qubit error rate as $0.98\%$, an average single-qubit $H$ error rate of $0.03\%$ and readout errors between $3\%$ and $7\%$.

\begin{figure}
\begin{subfigure}{0.24\textwidth}
    \centering
    \includegraphics[width=0.9\linewidth]{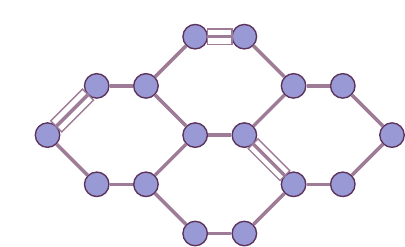}
    \caption{\footnotesize Hexagonal (IBM Washington)~\cite{Qiskit}}
    \label{fig:hex_arch}
\end{subfigure}
\begin{subfigure}{0.24\textwidth}
    \centering
    \includegraphics[width=0.9\linewidth]{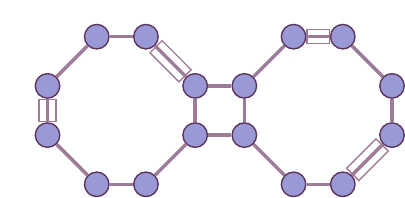}
    \caption{\footnotesize Octagonal (Rigetti Aspen)~\cite{rigetti_aspen}}
    \label{fig:oct_arch}
\end{subfigure}
\begin{subfigure}{0.24\textwidth}
    \centering
    \includegraphics[width=0.7\linewidth]{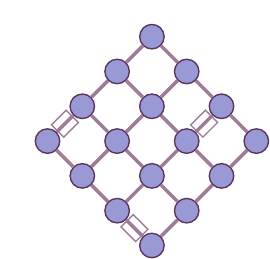}
    \caption{\footnotesize Grid (Google Sycamore)~\cite{sycamore}}
    \label{fig:squ_arch}
\end{subfigure}
\begin{subfigure}{0.24\textwidth}
    \centering
    \includegraphics[width=0.7\linewidth]{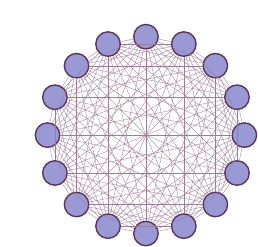}
    \caption{\footnotesize Fully Connected (IonQ Forte)~\cite{ionq}}
    \label{fig:ful_arch}
\end{subfigure}
\caption{\small Simulated device backends represent the coupling maps from a range of modern NISQ architectures. Devices are constructed where each qubit has a random state-dependent measurement error for both $\ket{0} \rightarrow \ket{1}$ and $\ket{1} \rightarrow \ket{0}$ in the range of $2\%$ to $8\%$. The boxed edges indicate an example patch on this device. \label{fig:simulated_device_backends}}
\end{figure}

\subsection{NISQ Device Benchmark Circuits}
Additionally, we benchmark the performance of these measurement error mitigation methods on IBM quantum devices using GHZ circuits. These represent the smallest full device entangling circuits, minimising the impact of gate errors while producing a non-classical distribution of measurement results. The GHZ circuits are constructed by performing a single-qubit Hadamard gate on qubit 0 of the device, and then performing a breadth first search of two-qubit CNOT operations across the coupling map. This construction ensures that there is no advantage gained by different qubit allocations, routing methods or other compiler optimisations.

\section{Evaluation} 		
\label{section:evaluation}
\subsection{Simulated Measurement Errors}

\begin{figure}
\centering
\begin{subfigure}[b]{0.49\textwidth}
\includegraphics[width=1\linewidth, trim=0cm 0cm 0cm 0.5cm]{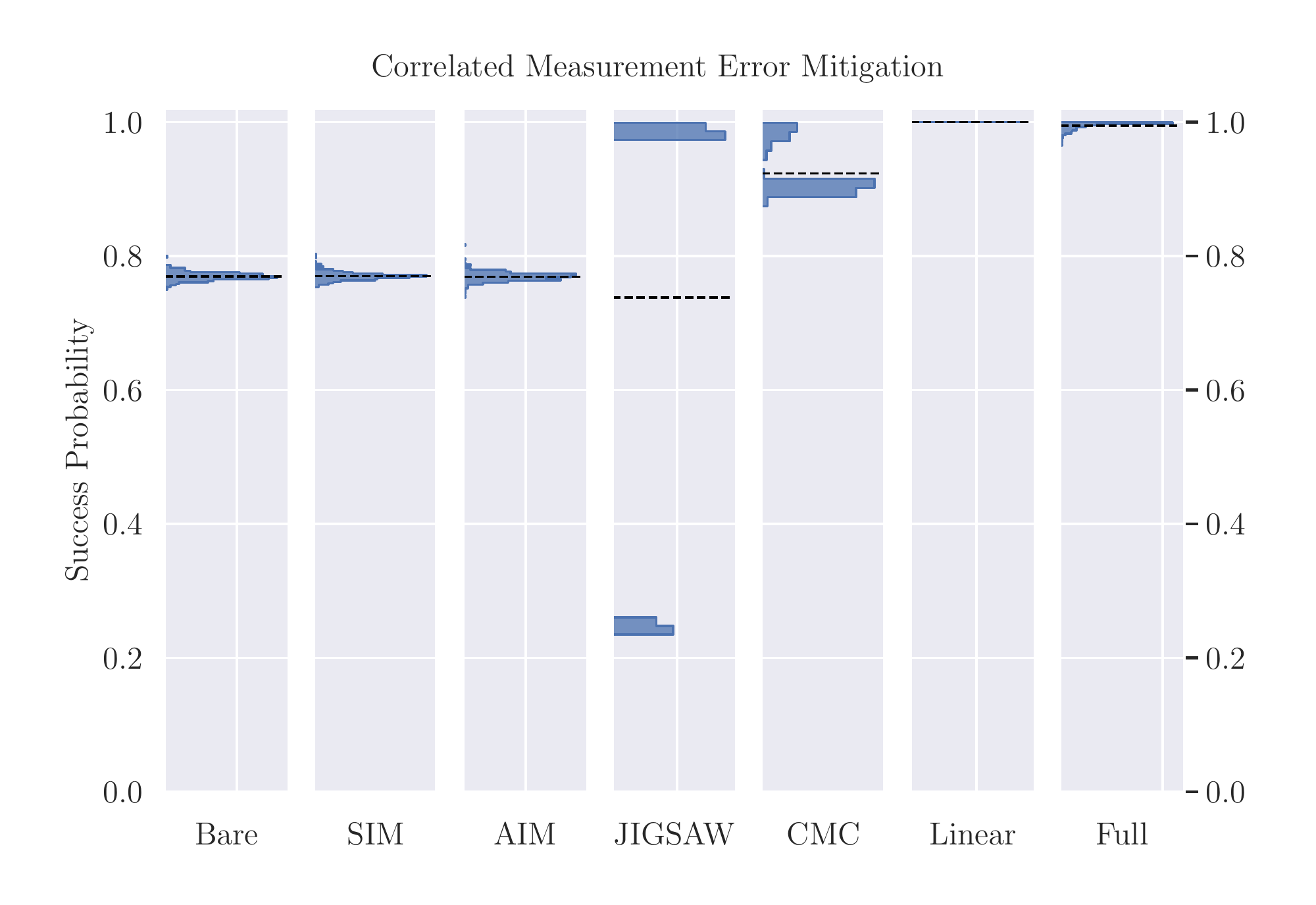}
\caption{\small Correlated Measurement Error Mitigation}
\end{subfigure}
\begin{subfigure}[b]{0.49\textwidth}
\includegraphics[width=1\linewidth, trim=0cm 0cm 0cm 0.5cm]{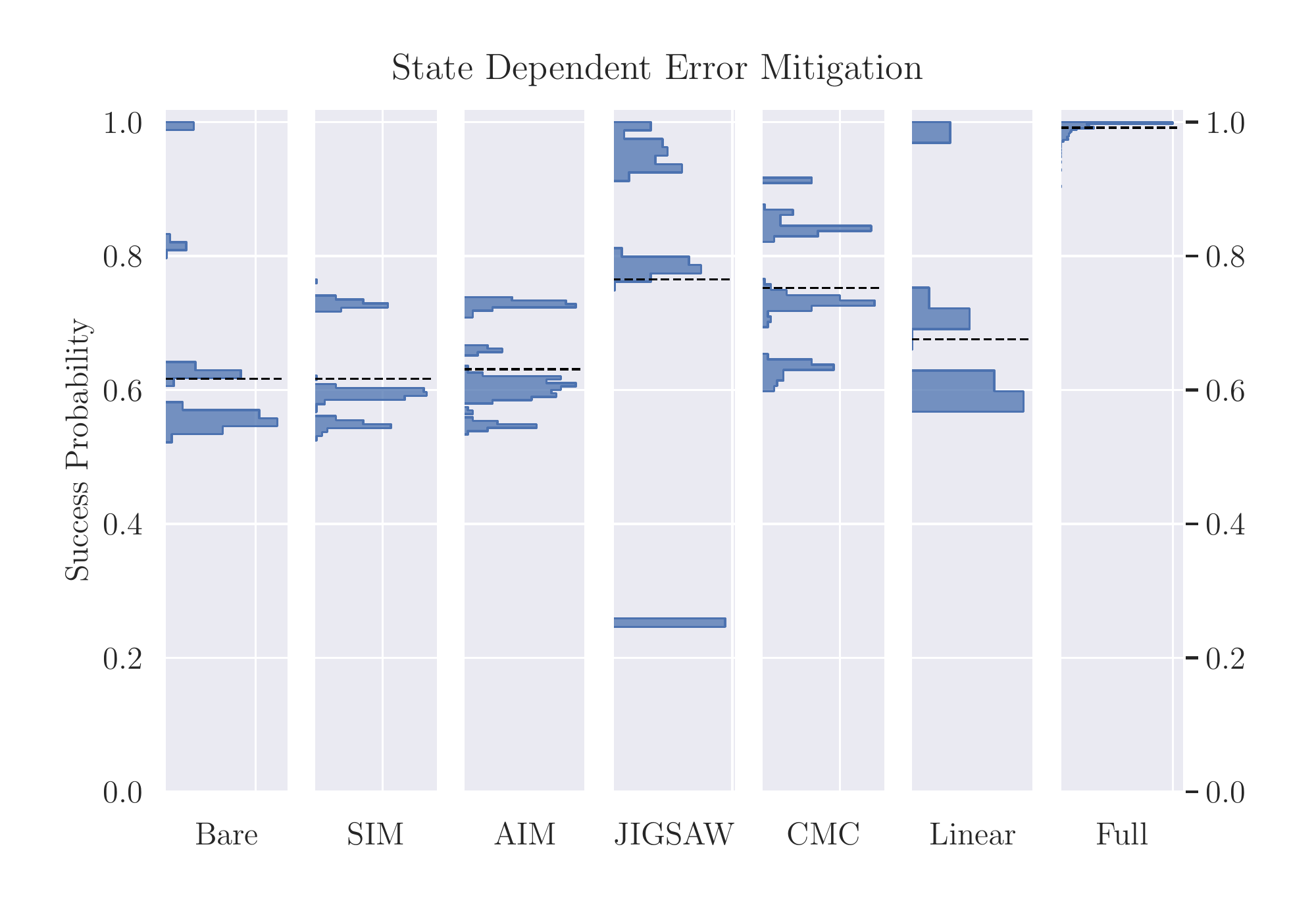}
\caption{\small State-Dependent Measurement Error Mitigation}
\end{subfigure}
\caption{\small (a) A sample correlated measurement error mitigation. (b) A sample state-dependent measurement error mitigation. Dashed lines indicate the average of each distribution. The `Bare' column shows the error distribution without before any mitigations are applied. The bifurcation of JIGSAW occurs where due to the pathological nature of the error model the sub-tables may be missing entries, leading to errors during renormalisation.\label{fig:sim_state_err}}
\vspace{-0.5cm}
\end{figure}
We benchmarked a range of measurement error mitigation methods over 136000 trials using a pair of known measurement error operations applied to the full set of $2^n$ computational basis states over four qubits, with results shown in Fig.~\ref{fig:sim_state_err}. Each mitigation method is afforded an equal number of measurements of the quantum system to perform any calibrations and operations of the circuit. As the classical input state is known, the success probability is the frequency that the classical output state is reported. `Bare' represents the distribution of errors without any mitigation. The correlated measurement error model applies only two qubit correlated errors. The state-dependent error model applies single qubit state-dependent errors, as a result the $\ket{0}^{\otimes n}$ state experiences no errors at all.  AIM and SIM perform their averaging techniques, which has no overall effect for correlated errors, and narrows the distribution of state-dependent errors. With these focused error models, JIGSAW suffers from an empty sub-table entry with high probability, and should not be considered representative of JIGSAW's performance on more rounded error models. CMC performs well in the presence of both classes of errors, but is outperformed by the Linear and Full methods, which require exponential overheads. As the total number of shots was constrained, the Full model suffers from slight sampling errors leading to a tail in its distribution.

\subsection{Simulated Architectures}

Using the architectures described in Fig.~\ref{fig:simulated_device_backends}, we consider the performance of our range of error mitigation techniques as a function of the topology and scale of the device. In each trial each method is provided with a fixed 16000 shots to reconstruct a GHZ state from the output of the simulated device. As the Qiskit statevector simulators limit readout errors on a per-qubit basis, the noise in these experiments is biased but not correlated.
\begin{figure}
%\centering
%\begin{center}
\hspace{-2em}
\includegraphics[width=1.1 \linewidth]{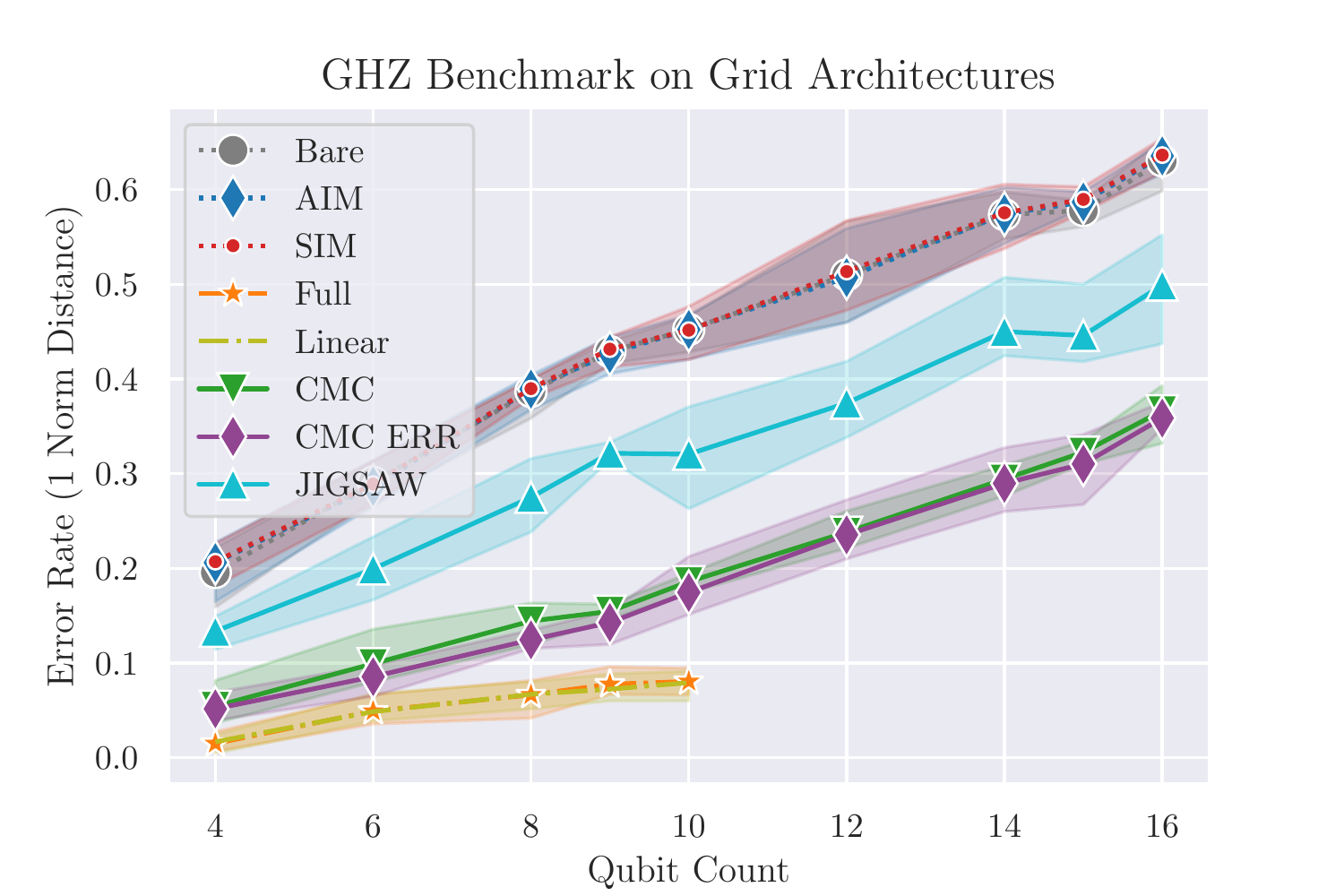}
%\end{center}
\caption{\small   Error rate of GHZ state preparation over a family of simulated devices with grid coupling maps as shown in Fig.~\ref{fig:squ_arch}. Each method is permitted 16000 shots with which to reconstruct a GHZ$_n$ state over $n$ qubits. \label{fig:grid_eval}}
\end{figure}

In Fig.~\ref{fig:grid_eval} and Fig.~\ref{fig:hex_eval} that CMC and CMC-ERR performs the best of the non-exponential methods over grid and hexagonal architectures. The Full and Linear methods provide the greatest reduction in one-norm distance, but require exponential overheads which limits their application to low qubit counts. AIM and SIM are nearly indistinguishable from the bare error rate. JIGSAW outperforms the averaging methods, but is in turn outperformed by CMC. As the statevector simulator's measurement errors are applied on a per-qubit basis, this guarantees the linearity of the error model. As a result the linear calibration performs as well as the full calibration over these simulations up to the difference in sampling during characterisation.
\begin{figure}
%\centering
%\begin{center}
\hspace{-2em}
\includegraphics[width=1.1 \linewidth]{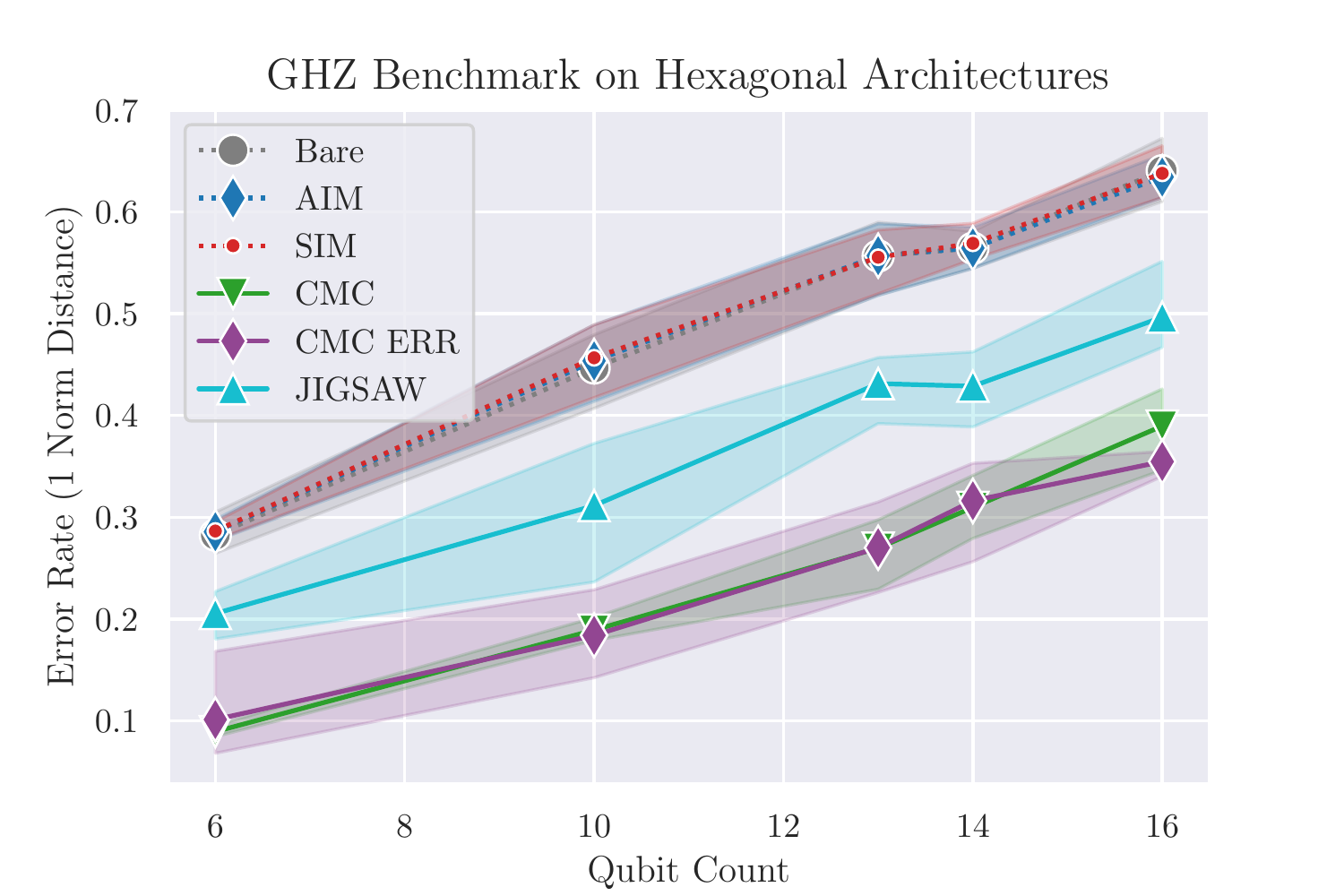}
%\end{center}
\caption{\small Error rate of GHZ state preparation over a family of simulated devices with grid coupling maps as shown in Fig.~\ref{fig:hex_arch} Each method is permitted 16000 shots with which to reconstruct a GHZ$_n$ state over $n$ qubits.\label{fig:hex_eval}}
\end{figure}
In the case of the fully connected topology seen in Fig.~\ref{fig:ful_arch}, the number of edges scales quadratically with the number of qubits. As a result the CMC method begins to suffer from a reduced number of shots per coupling map patch, which significantly reduces the accuracy of the method, as can be seen in Fig.~\ref{fig:full_eval}. For this dense coupling map JIGSAW slightly outperforms CMC, and would be expected to significantly outperform it for larger devices. CMC-ERR outperforms both CMC and JIGSAW, using the constraint of $n$ edges to mitigate the quadratic scaling issues of base CMC.

\begin{figure}
%\centering
%\begin{center}
\hspace{-2em}
\includegraphics[width=1.1 \linewidth]{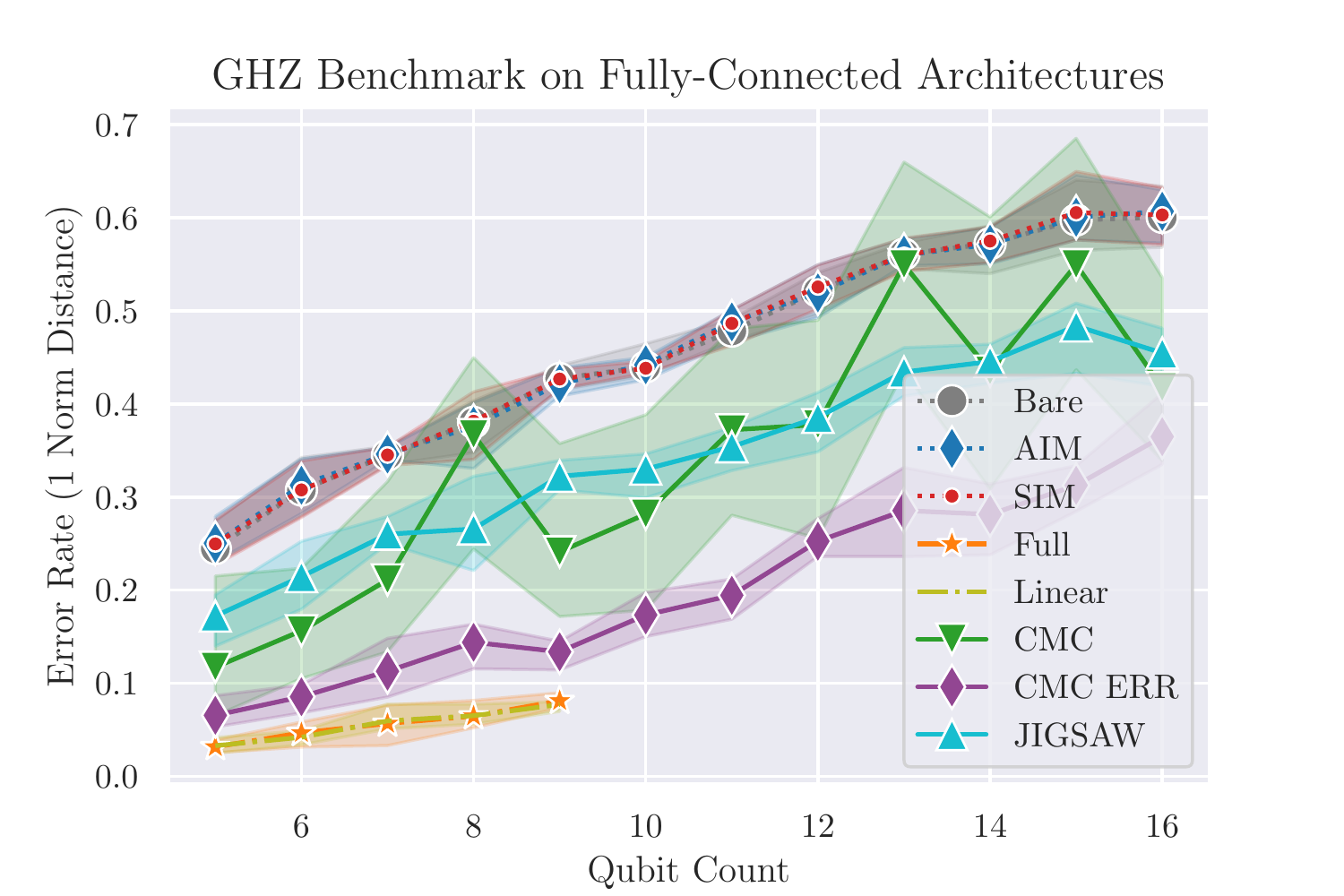}
%\end{center}
\caption{\small Error rate of GHZ state preparation over a family of simulated devices with grid coupling maps as shown in Fig.~\ref{fig:ful_arch}  Each method is permitted 16000 shots with which to reconstruct a GHZ$_n$ state over $n$ qubits. This poses some problems for the CMC method as the number of edges for the complete coupling map scales quadratically with the number of qubits.\label{fig:full_eval}}
\end{figure}

For completeness, we include the octagonal topology with the same error model as the other families of devices. At 16 qubits, JIGSAW achieves a $23\%$ reduction over the baseline error rate, while $CMC$ reduces the error rate by $37\%$. For the same octagonal device, AIM and SIM are within $1\%$ of the initial error rate.

\subsection{NISQ Device Benchmarks}

\begin{table}[t!]\small
\centering
\begin{tabular}{lcccc}
        \toprule
    \multirow{2}*{\bf Method} & \multicolumn{4}{c}{GHZ Circuit Benchmarks}  \\
    & {\footnotesize Manila - $5$} & {\footnotesize Lima - $5$} &{\footnotesize Quito - $5$} & {\footnotesize Nairobi - $7$} \\
    \midrule
    Bare & $0.20\substack{+0.10 \\ -0.04}$ & $0.23\substack{+0.02 \\ -0.01}$ & $0.26\substack{+0.01 \\ -0.01}$ & $0.56\substack{+0.02 \\ -0.01}$ \\
    Full & $0.10\substack{+0.11 \\ -0.04}$ & $0.07\substack{+0.02 \\ -0.01}$ & $0.04\substack{+0.02 \\ -0.01}$ & N/A\\
    Linear & $0.11\substack{+0.02 \\ -0.02}$ & $0.06\substack{+0.02 \\ -0.01}$ & $0.08\substack{+0.02 \\ -0.02}$ & N/A \\
    AIM & $0.18\substack{+0.03 \\ -0.02}$ & $0.23\substack{+0.01 \\ -0.01}$ & $0.26\substack{+0.01 \\ -0.01}$ & $0.57\substack{+0.01 \\ -0.01}$\\
    SIM & $0.19\substack{+0.03 \\ -0.02}$ & $0.23\substack{+0.01 \\ -0.01}$ &  $0.27\substack{+0.01 \\ -0.01}$ & $0.62\substack{+0.01 \\ -0.02}$\\
    JIGSAW & $0.18\substack{+0.06 \\ -0.06}$ & ${\bf 0.17\substack{\bf +0.06 \\ \bf -0.05}}$ & $0.21\substack{+0.04 \\ -0.04}$ & $0.52\substack{+0.19 \\ -0.23}$\\
    {\it CMC } & $0.14\substack{+0.09 \\ -0.05}$ & ${\bf 0.17\substack{\bf +0.01 \\ \bf -0.04}}$ & ${\bf 0.16\substack{\bf +0.01 \\ \bf -0.01}}$ & $0.64\substack{+0.02 \\ -0.06}$\\
    {\it $\text{CMC}_{\text{ERR}}$} & ${\bf0.13\substack{\bf +0.08 \\ \bf -0.03}}$ & $0.25\substack{+0.03 \\ -0.03}$ & $0.17\substack{+0.10 \\ -0.03}$ & ${\bf 0.33\substack{\bf +0.41 \\ \bf -0.12}}$\\
    \bottomrule
\end{tabular}
\caption{\small GHZ benchmarks of the 1-norm distance between output distribution and ideal $\ket{+}^{\otimes n}$ state. The relative performance of CMC and $\text{CMC}_{\text{ERR}}$ depends on the `uniformity' of the error distributions for that device as seen in Fig.~\ref{fig:corr_coupling_demo}\label{table:GHZ_dev} The best non-exponential method is bolded in each column.}
\end{table}

Lastly we compare the performance of these methods on physical quantum devices. IBM Manila and Nairobi have local but non-coupling map aligned correlated errors, while Lima and Quito have locally uniform error profiles, as seen in Fig.~\ref{fig:corr_coupling_demo}. From this we would expect a relative advantage for CMC-ERR on Manila and Nairobi, and for CMC on Lima and Quito. This is supported by the results in Table ~\ref{table:GHZ_dev}. Each method is allocated 32000 shots to perform both calibration and any required circuit executions. For all five qubit devices exponential methods achieve the best performance. However, at the seven qubit mark these methods begin to encounter scaling issues, with the Full calibration approach exceeding 100 calibration circuits. Of the non-exponential methods CMC and CMC-ERR beat or match JIGSAW on all devices. 

These results reveal a strong association on the performance of each method with the underlying error model of the device. Correlated errors on IBMQ-Nairobi are almost anti-aligned with the device's coupling map, as seen in Fig.~\ref{fig:err_map_demo}. This explains the poor performance of base CMC on this device which failed to characterise most of the correlated errors. Similarly JIGSAW, which relies on calibrating against random pairs of qubits, performs best on relatively even error distributions. In several instances JIGSAW's best case matched CMC or CMC-ERR's best case, but had a worse average performance due to its reliance on the randomised calibration pairs. CMC-ERR exhibited a $41\%$ reduction in the average error rate on IBMQ Nairobi, demonstrating the efficacy of tailoring correlated error mitigation methods to device noise profiles.

\section{Discussion}        
\label{section:discussion}

\subsection{Scalability}
The IBMQ implementations of the Full and Linear calibration matrices quickly encounter scaling issues. These methods require the construction of a $2^n\times2^n$ matrix. The Full method performs this using $2^n$ calibration circuits. For $n > 10$ it becomes unfeasible to queue and execute all the required calibration circuits. Alongside the constraints of constructing this matrix we must also consider the classical computational overheads. For example at $n = 14$ the dense calibration matrix (using 32 bit floating point representation) occupies 32GB of RAM. Calculating the inverse of this matrix and applying it to the sparse output vector of the measurement results is computationally unfeasible. By comparison using a na\"ive COO sparse matrix representation with CMC, 32GB affords us 32 qubits. More  memory efficient sparse matrix constructions may be made given the regular structure of the CMC matrices. 

For calibration matrix methods (Full, Linear and CMC), as long as the error profile of the device does not drift significantly, it is possible to apply the same calibration matrix to multiple circuits executed on the same device. Circuit dependent methods (AIM, SIM, and JIGSAW) must be re-run from scratch for every new circuit that is executed. We find that ERR characterisations are stable for a given device on the order of weeks between significant recalibrations.

\subsection{Linearity of Edge Counts on NISQ Devices}
The main concern for the scalability of CMC is the number of edges in the coupling map. A non-sparse coupling map increases the number of two qubit calibrations that are performed and decreases the number of calibration patches that may be performed simultaneously. Table~\ref{table:edge_count} shows the number of edges as a function of the number of qubits for a range of modern architectures.
\begin{table}[h]
    \centering
    \begin{tabular}{lcr}
        \toprule
        Architecture & Example & Edge Count$(n)$\\
        \midrule
        {\footnotesize Linear} & {\footnotesize Honeywell H1}{\footnotesize~\cite{honeywell}} & {\footnotesize $n - 1$} \\
        {\footnotesize Grid}{(\footnotesize Fig.~\ref{fig:squ_arch})} & {\footnotesize Google Sycamore}{\footnotesize~\cite{sycamore}} & {\footnotesize $2n + c + r$}\\
        {\footnotesize Local Grid}{(\footnotesize Fig.~\ref{fig:coupling_tokyo})} & {\footnotesize IBM Tokyo }{\footnotesize~\cite{Qiskit}} & {\footnotesize $4n + c + r$}\\
        {\footnotesize Hexagonal}{(\footnotesize Fig.~\ref{fig:hex_arch})} & {\footnotesize Rigetti ACORN}{\footnotesize \cite{rigetti}} & {\footnotesize $(n - 1) + cr$}\\
        {\footnotesize Heavy Hex}{(\footnotesize Fig.~\ref{fig:hex_arch})} & {\footnotesize IBM Washington}{\footnotesize \cite{Qiskit}} & {\footnotesize $(n - 1) + cr$}\\
        {\footnotesize Octagonal}{(\footnotesize Fig.~\ref{fig:oct_arch})} & {\footnotesize Rigetti ASPEN}{\footnotesize \cite{rigetti_aspen}} & {\footnotesize $\frac{3n}{2} - 2r - 2c$}\\
        {\footnotesize Fully Connected}{(\footnotesize Fig.~\ref{fig:ful_arch})} & {\footnotesize IonQ Forte}{\footnotesize \cite{ionq}} &  {\footnotesize $\frac{1}{2}(n^2 - n)$}\\
        \bottomrule
    \end{tabular}
    \caption{\small Edge count as a function of the number of qubits $n$, rows $r$ and columns $c$ for a range of device architectures. For all architectures $n \ge rc$. All architectures excepting IonQ's grow the number of edges linearly with the number of qubits.\label{table:edge_count}}
\end{table}

IonQ's fully-connected devices~\cite{ionq}(Fig.~\ref{fig:ful_arch}) are the only family of architectures with a greater than linear growth in the number of edges, hence it is not scalable to perform bare CMC over this fully connected graph. The construction of CMC-ERR avoids this scaling problem bounds that the total number of calibration patches by the number of qubits.

\section{Conclusion} 		
\label{section:conclusion}
In this paper we have demonstrated CMC and CMC-ERR, which are novel methods for efficiently constructing and joining sparse and scalable measurement calibration matrices. These methods do not increase the number of measurement shots on the device. Results on IBMQ devices have achieved up to a $41\%$ reduction in the error rate rate of the circuit, averaging at $35\%$ over the experiments performed and outperform all other non-exponential methods. Our results demonstrate a strong association between the performance of measurement error mitigation methods with the underlying error model of the device. 

All code required to replicate these results can be found in the corresponding \href{https://github.com/Alan-Robertson/quantum_measurement_error_mitigation}{git repository}.

%\section*{Acknowledgements}
%%%%%%% -- PAPER CONTENT ENDS -- %%%%%%%%

%%%%%%%%% -- BIB STYLE AND FILE -- %%%%%%%%
\bibliographystyle{IEEEtranS}
\bibliography{refs}

% Generated by IEEEtranS.bst, version: 1.13 (2008/09/30)
\begin{thebibliography}{10}
\providecommand{\url}[1]{#1}
\csname url@samestyle\endcsname
\providecommand{\newblock}{\relax}
\providecommand{\bibinfo}[2]{#2}
\providecommand{\BIBentrySTDinterwordspacing}{\spaceskip=0pt\relax}
\providecommand{\BIBentryALTinterwordstretchfactor}{4}
\providecommand{\BIBentryALTinterwordspacing}{\spaceskip=\fontdimen2\font plus
\BIBentryALTinterwordstretchfactor\fontdimen3\font minus
  \fontdimen4\font\relax}
\providecommand{\BIBforeignlanguage}[2]{{%
\expandafter\ifx\csname l@#1\endcsname\relax
\typeout{** WARNING: IEEEtranS.bst: No hyphenation pattern has been}%
\typeout{** loaded for the language `#1'. Using the pattern for}%
\typeout{** the default language instead.}%
\else
\language=\csname l@#1\endcsname
\fi
#2}}
\providecommand{\BIBdecl}{\relax}
\BIBdecl

\bibitem{Qiskit}
\BIBentryALTinterwordspacing
G.~Aleksandrowicz, T.~Alexander, P.~Barkoutsos, L.~Bello, Y.~Ben-Haim,
  D.~Bucher, F.~J. Cabrera-Hernández, J.~Carballo-Franquis, A.~Chen, C.-F.
  Chen, J.~M. Chow, A.~D. Córcoles-Gonzales, A.~J. Cross, A.~Cross,
  J.~Cruz-Benito, C.~Culver, S.~D. L.~P. González, E.~D.~L. Torre, D.~Ding,
  E.~Dumitrescu, I.~Duran, P.~Eendebak, M.~Everitt, I.~F. Sertage, A.~Frisch,
  A.~Fuhrer, J.~Gambetta, B.~G. Gago, J.~Gomez-Mosquera, D.~Greenberg,
  I.~Hamamura, V.~Havlicek, J.~Hellmers, Łukasz Herok, H.~Horii, S.~Hu,
  T.~Imamichi, T.~Itoko, A.~Javadi-Abhari, N.~Kanazawa, A.~Karazeev,
  K.~Krsulich, P.~Liu, Y.~Luh, Y.~Maeng, M.~Marques, F.~J. Martín-Fernández,
  D.~T. McClure, D.~McKay, S.~Meesala, A.~Mezzacapo, N.~Moll, D.~M. Rodríguez,
  G.~Nannicini, P.~Nation, P.~Ollitrault, L.~J. O'Riordan, H.~Paik, J.~Pérez,
  A.~Phan, M.~Pistoia, V.~Prutyanov, M.~Reuter, J.~Rice, A.~R. Davila, R.~H.~P.
  Rudy, M.~Ryu, N.~Sathaye, C.~Schnabel, E.~Schoute, K.~Setia, Y.~Shi,
  A.~Silva, Y.~Siraichi, S.~Sivarajah, J.~A. Smolin, M.~Soeken, H.~Takahashi,
  I.~Tavernelli, C.~Taylor, P.~Taylour, K.~Trabing, M.~Treinish, W.~Turner,
  D.~Vogt-Lee, C.~Vuillot, J.~A. Wildstrom, J.~Wilson, E.~Winston, C.~Wood,
  S.~Wood, S.~Wörner, I.~Y. Akhalwaya, and C.~Zoufal, ``{Qiskit: An
  Open-source Framework for Quantum Computing},'' Jan. 2019. [Online].
  Available: \url{https://doi.org/10.5281/zenodo.2562111}
\BIBentrySTDinterwordspacing

\bibitem{full_stack}
\BIBentryALTinterwordspacing
Y.~Alexeev, D.~Bacon, K.~R. Brown, R.~Calderbank, L.~D. Carr, F.~T. Chong,
  B.~DeMarco, D.~Englund, E.~Farhi, B.~Fefferman, A.~V. Gorshkov, A.~Houck,
  J.~Kim, S.~Kimmel, M.~Lange, S.~Lloyd, M.~D. Lukin, D.~Maslov, P.~Maunz,
  C.~Monroe, J.~Preskill, M.~Roetteler, M.~J. Savage, and J.~Thompson,
  ``Quantum computer systems for scientific discovery,'' \emph{PRX Quantum},
  vol.~2, p. 017001, Feb 2021. [Online]. Available:
  \url{https://link.aps.org/doi/10.1103/PRXQuantum.2.017001}
\BIBentrySTDinterwordspacing

\bibitem{altepeter}
J.~Altepeter, D.~James, and P.~Kwiat, ``Quantum state tomography,'' 2017.

\bibitem{sycamore}
\BIBentryALTinterwordspacing
F.~Arute, K.~Arya, R.~Babbush, D.~Bacon, J.~C. Bardin, R.~Barends, R.~Biswas,
  S.~Boixo, F.~G. S.~L. Brandao, D.~A. Buell, B.~Burkett, Y.~Chen, Z.~Chen,
  B.~Chiaro, R.~Collins, W.~Courtney, A.~Dunsworth, E.~Farhi, B.~Foxen,
  A.~Fowler, C.~Gidney, M.~Giustina, R.~Graff, K.~Guerin, S.~Habegger, M.~P.
  Harrigan, M.~J. Hartmann, A.~Ho, M.~Hoffmann, T.~Huang, T.~S. Humble, S.~V.
  Isakov, E.~Jeffrey, Z.~Jiang, D.~Kafri, K.~Kechedzhi, J.~Kelly, P.~V. Klimov,
  S.~Knysh, A.~Korotkov, F.~Kostritsa, D.~Landhuis, M.~Lindmark, E.~Lucero,
  D.~Lyakh, S.~Mandr{\`{a}}, J.~R. McClean, M.~McEwen, A.~Megrant, X.~Mi,
  K.~Michielsen, M.~Mohseni, J.~Mutus, O.~Naaman, M.~Neeley, C.~Neill, M.~Y.
  Niu, E.~Ostby, A.~Petukhov, J.~C. Platt, C.~Quintana, E.~G. Rieffel,
  P.~Roushan, N.~C. Rubin, D.~Sank, K.~J. Satzinger, V.~Smelyanskiy, K.~J.
  Sung, M.~D. Trevithick, A.~Vainsencher, B.~Villalonga, T.~White, Z.~J. Yao,
  P.~Yeh, A.~Zalcman, H.~Neven, and J.~M. Martinis, ``Quantum supremacy using a
  programmable superconducting processor,'' \emph{Nature}, vol. 574, no. 7779,
  pp. 505--510, oct 2019. [Online]. Available:
  \url{https://doi.org/10.1038%2Fs41586-019-1666-5}
\BIBentrySTDinterwordspacing

\bibitem{berry}
\BIBentryALTinterwordspacing
D.~W. Berry, B.~L. Higgins, S.~D. Bartlett, M.~W. Mitchell, G.~J. Pryde, and
  H.~M. Wiseman, ``How to perform the most accurate possible phase
  measurements,'' \emph{Phys. Rev. A}, vol.~80, p. 052114, Nov 2009. [Online].
  Available: \url{https://link.aps.org/doi/10.1103/PhysRevA.80.052114}
\BIBentrySTDinterwordspacing

\bibitem{wildcard_rbk}
\BIBentryALTinterwordspacing
R.~Blume-Kohout, K.~Rudinger, E.~Nielsen, T.~Proctor, and K.~Young, ``Wildcard
  error: Quantifying unmodeled errors in quantum processors,'' 2020. [Online].
  Available: \url{https://arxiv.org/abs/2012.12231}
\BIBentrySTDinterwordspacing

\bibitem{jigsaw}
\BIBentryALTinterwordspacing
P.~Das, S.~Tannu, and M.~Qureshi, ``Jigsaw: Boosting fidelity of nisq programs
  via measurement subsetting,'' in \emph{MICRO-54: 54th Annual IEEE/ACM
  International Symposium on Microarchitecture}, ser. MICRO '21.\hskip 1em plus
  0.5em minus 0.4em\relax New York, NY, USA: Association for Computing
  Machinery, 2021, p. 937–949. [Online]. Available:
  \url{https://doi.org/10.1145/3466752.3480044}
\BIBentrySTDinterwordspacing

\bibitem{deutsch_rapid_1992}
D.~Deutsch and R.~Jozsa, ``\BIBforeignlanguage{en}{Rapid {{Solution}} of
  {{Problems}} by {{Quantum Computation}}},''
  \emph{\BIBforeignlanguage{en}{Proceedings of the Royal Society of London A:
  Mathematical, Physical and Engineering Sciences}}, vol. 439, no. 1907, pp.
  553--558, Dec. 1992.

\bibitem{gottesman}
\BIBentryALTinterwordspacing
D.~Gottesman, ``An introduction to quantum error correction and fault-tolerant
  quantum computation,'' \emph{quant-ph}. [Online]. Available:
  \url{arXiv:0904.2557v1}
\BIBentrySTDinterwordspacing

\bibitem{qinfer-1_0}
\BIBentryALTinterwordspacing
C.~Granade, C.~Ferrie, S.~Casagrande, I.~Hincks, M.~Kononenko, T.~Alexander,
  and Y.~Sanders, ``{QInfer}: Library for statistical inference in quantum
  information,'' Sep. 2016. [Online]. Available:
  \url{http://dx.doi.org/10.5281/zenodo.157007}
\BIBentrySTDinterwordspacing

\bibitem{Granade_2017}
\BIBentryALTinterwordspacing
C.~Granade, C.~Ferrie, and S.~T. Flammia, ``Practical adaptive quantum
  tomography,'' \emph{New Journal of Physics}, vol.~19, no.~11, p. 113017, nov
  2017. [Online]. Available: \url{https://doi.org/10.1088%2F1367-2630%2Faa8fe6}
\BIBentrySTDinterwordspacing

\bibitem{greenbaum_tomography}
\BIBentryALTinterwordspacing
D.~Greenbaum, ``Introduction to quantum gate set tomography,'' 2015. [Online].
  Available: \url{https://arxiv.org/abs/1509.02921}
\BIBentrySTDinterwordspacing

\bibitem{grover_fast_1996}
L.~K. Grover, ``A fast quantum mechanical algorithm for database search,''
  \emph{arXiv:quant-ph/9605043}, May 1996.

\bibitem{riddhi}
R.~S. Gupta, A.~R. Milne, C.~L. Edmunds, C.~Hempel, and M.~J. Biercuk,
  ``Autonomous adaptive noise characterization in quantum computers,''
  \emph{npj Quantum Inf}, vol.~6, 2020.

\bibitem{harper_correlated}
R.~Harper, S.~T. Flammia, and J.~J. Wallman, ``Efficient learning of quantum
  noise,'' \emph{Nat. Phys}, 2020.

\bibitem{harper_fault-tolerant_2019}
\BIBentryALTinterwordspacing
R.~Harper and S.~T. Flammia, ``Fault-{Tolerant} {Logical} {Gates} in the {IBM}
  {Quantum} {Experience},'' \emph{Physical Review Letters}, vol. 122, no.~8, p.
  080504, Feb. 2019. [Online]. Available:
  \url{https://link.aps.org/doi/10.1103/PhysRevLett.122.080504}
\BIBentrySTDinterwordspacing

\bibitem{howard}
\BIBentryALTinterwordspacing
M.~Howard, J.~Twamley, C.~Wittmann, T.~Gaebel, F.~Jelezko, and J.~Wrachtrup,
  ``Quantum process tomography and linblad estimation of a solid-state qubit,''
  \emph{New Journal of Physics}, vol.~8, no.~3, pp. 33--33, mar 2006. [Online].
  Available: \url{https://doi.org/10.1088%2F1367-2630%2F8%2F3%2F033}
\BIBentrySTDinterwordspacing

\bibitem{kassal_polynomial-time_2008}
I.~Kassal, S.~P. Jordan, P.~J. Love, M.~Mohseni, and A.~Aspuru-Guzik,
  ``\BIBforeignlanguage{en}{Polynomial-time quantum algorithm for the
  simulation of chemical dynamics},'' \emph{\BIBforeignlanguage{en}{Proceedings
  of the National Academy of Sciences}}, vol. 105, no.~48, pp.
  18\,681--18\,686, Feb. 2008.

\bibitem{rigetti_aspen}
\BIBentryALTinterwordspacing
A.~C.~Y. Li, M.~S. Alam, T.~Iadecola, A.~Jahin, D.~M. Kurkcuoglu, R.~Li, P.~P.
  Orth, A.~B. Özgüler, G.~N. Perdue, and N.~M. Tubman, ``Benchmarking
  variational quantum eigensolvers for the square-octagon-lattice kitaev
  model,'' 2021. [Online]. Available: \url{https://arxiv.org/abs/2108.13375}
\BIBentrySTDinterwordspacing

\bibitem{emerson_rb}
\BIBentryALTinterwordspacing
E.~Magesan, J.~M. Gambetta, and J.~Emerson, ``Scalable and robust randomized
  benchmarking of quantum processes,'' \emph{Phys. Rev. Lett.}, vol. 106, p.
  180504, May 2011. [Online]. Available:
  \url{https://link.aps.org/doi/10.1103/PhysRevLett.106.180504}
\BIBentrySTDinterwordspacing

\bibitem{magesan_characterizing_2012}
E.~Magesan, J.~M. Gambetta, and J.~Emerson, ``Characterizing {{Quantum Gates}}
  via {{Randomized Benchmarking}},'' \emph{Physical Review A}, vol.~85, no.~4,
  Apr. 2012.

\bibitem{merkel_tomography}
\BIBentryALTinterwordspacing
S.~T. Merkel, J.~M. Gambetta, J.~A. Smolin, S.~Poletto, A.~D. C\'orcoles, B.~R.
  Johnson, C.~A. Ryan, and M.~Steffen, ``Self-consistent quantum process
  tomography,'' \emph{Phys. Rev. A}, vol.~87, p. 062119, Jun 2013. [Online].
  Available: \url{https://link.aps.org/doi/10.1103/PhysRevA.87.062119}
\BIBentrySTDinterwordspacing

\bibitem{martonosi_crosstalk}
\BIBentryALTinterwordspacing
P.~Murali, D.~C. Mckay, M.~Martonosi, and A.~Javadi-Abhari, ``Software
  mitigation of crosstalk on noisy intermediate-scale quantum computers,'' in
  \emph{Proceedings of the Twenty-Fifth International Conference on
  Architectural Support for Programming Languages and Operating Systems}, ser.
  ASPLOS '20.\hskip 1em plus 0.5em minus 0.4em\relax New York, NY, USA:
  Association for Computing Machinery, 2020, p. 1001–1016. [Online].
  Available: \url{https://doi.org/10.1145/3373376.3378477}
\BIBentrySTDinterwordspacing

\bibitem{nielsen}
M.~Nielsen and I.~Chuang, \emph{Quantum Computation and Quantum
  Information}.\hskip 1em plus 0.5em minus 0.4em\relax Cambridge University
  Press.

\bibitem{sc_estimating}
T.~{Patel} and D.~{Tiwari}, ``Veritas: Accurately estimating the correct output
  on noisy intermediate-scale quantum computers,'' in \emph{SC20: International
  Conference for High Performance Computing, Networking, Storage and Analysis},
  2020, pp. 1--16.

\bibitem{honeywell}
\BIBentryALTinterwordspacing
J.~M. Pino, J.~M. Dreiling, C.~Figgatt, J.~P. Gaebler, S.~A. Moses, M.~S.
  Allman, C.~H. Baldwin, M.~Foss-Feig, D.~Hayes, K.~Mayer, C.~Ryan-Anderson,
  and B.~Neyenhuis, ``Demonstration of the trapped-ion quantum {CCD} computer
  architecture,'' \emph{Nature}, vol. 592, no. 7853, pp. 209--213, apr 2021.
  [Online]. Available: \url{https://doi.org/10.1038%2Fs41586-021-03318-4}
\BIBentrySTDinterwordspacing

\bibitem{preskill_nisq}
\BIBentryALTinterwordspacing
J.~Preskill, ``Quantum {C}omputing in the {NISQ} era and beyond,''
  \emph{{Quantum}}, vol.~2, p.~79, Aug. 2018. [Online]. Available:
  \url{https://doi.org/10.22331/q-2018-08-06-79}
\BIBentrySTDinterwordspacing

\bibitem{Proctor_2021}
\BIBentryALTinterwordspacing
T.~Proctor, K.~Rudinger, K.~Young, E.~Nielsen, and R.~Blume-Kohout, ``Measuring
  the capabilities of quantum computers,'' \emph{Nature Physics}, vol.~18,
  no.~1, pp. 75--79, dec 2021. [Online]. Available:
  \url{https://doi.org/10.1038%2Fs41567-021-01409-7}
\BIBentrySTDinterwordspacing

\bibitem{shor_polynomial-time_1995}
P.~W. Shor, ``Polynomial-{{Time Algorithms}} for {{Prime Factorization}} and
  {{Discrete Logarithms}} on a {{Quantum Computer}},''
  \emph{arXiv:quant-ph/9508027}, Aug. 1995.

\bibitem{rigetti}
\BIBentryALTinterwordspacing
R.~S. Smith, M.~J. Curtis, and W.~J. Zeng, ``A practical quantum instruction
  set architecture,'' 2016. [Online]. Available:
  \url{https://arxiv.org/abs/1608.03355}
\BIBentrySTDinterwordspacing

\bibitem{swamit_state_dep}
S.~Tannu and M.~Qureshi, ``Mitigating measurement errors in quantum computers
  by exploiting state-dependent bias,'' \emph{MICRO}, 2019.

\bibitem{unruh}
W.~Unruh, ``Maintaining coherence in quantum computers,'' vol.~51.

\bibitem{wood_open_quantum}
\BIBentryALTinterwordspacing
C.~J. Wood, J.~D. Biamonte, and D.~G. Cory, ``Tensor networks and graphical
  calculus for open quantum systems,'' 2011. [Online]. Available:
  \url{https://arxiv.org/abs/1111.6950}
\BIBentrySTDinterwordspacing

\bibitem{wooters}
W.~Wooters and W.~Zurek, ``A single quantum cannot be cloned,'' \emph{Nature},
  no. 299, pp. 802--803.

\bibitem{ionq}
\BIBentryALTinterwordspacing
K.~Wright, K.~M. Beck, S.~Debnath, J.~M. Amini, Y.~Nam, N.~Grzesiak, J.-S.
  Chen, N.~C. Pisenti, M.~Chmielewski, C.~Collins, K.~M. Hudek, J.~Mizrahi,
  J.~D. Wong-Campos, S.~Allen, J.~Apisdorf, P.~Solomon, M.~Williams, A.~M.
  Ducore, A.~Blinov, S.~M. Kreikemeier, V.~Chaplin, M.~Keesan, C.~Monroe, and
  J.~Kim, ``Benchmarking an 11-qubit quantum computer,'' \emph{Nature
  Communications}, vol.~10, no.~1, nov 2019. [Online]. Available:
  \url{https://doi.org/10.1038%2Fs41467-019-13534-2}
\BIBentrySTDinterwordspacing

\end{thebibliography}
%%%%%%%%%%%%%%%%%%%%%%%%%%%%%%%%%%%%

\end{document}